\documentclass[sigconf]{acmart}

\usepackage{color}
\usepackage{pifont}
\usepackage{subfigure} 
\usepackage{float}

\usepackage{booktabs}
\usepackage{multirow}

\usepackage[toc,page]{appendix} 
\usepackage{tabularx}
\usepackage{url}
\usepackage{enumitem}
\setlist[itemize]{leftmargin=*}

\newcommand{\ms}[1]{{\color{black} #1}}

\usepackage{color, colortbl}
\usepackage{listings}
\usepackage[ruled]{algorithm2e}

\definecolor{color1}{HTML}{CAEEFB}
\definecolor{color2}{HTML}{D9F2D0}
\definecolor{color3}{HTML}{FBE3D6}
\definecolor{darkblue}{RGB}{31,78,121}

\definecolor{orange}{RGB}{255,69,0}
\AtBeginDocument{%
  \providecommand\BibTeX{{%
    \normalfont B\kern-0.5em{\scshape i\kern-0.25em b}\kern-0.8em\TeX}}}

\copyrightyear{2025}
\acmYear{2025}
\setcopyright{acmlicensed}\acmConference[CHI '25]{CHI Conference on Human Factors in Computing Systems}{April 26-May 1, 2025}{Yokohama, Japan}
\acmBooktitle{CHI Conference on Human Factors in Computing Systems (CHI '25), April 26-May 1, 2025, Yokohama, Japan}
\acmDOI{10.1145/3706598.3713748}
\acmISBN{979-8-4007-1394-1/25/04}

\begin{document}




\title{DBox: Scaffolding Algorithmic Programming Learning through Learner-LLM Co-Decomposition}






\author{Shuai Ma}
\authornote{Work done during the first author's PhD studies at The Hong Kong University of Science and Technology.}
\orcid{0000-0002-7658-292X}
\affiliation{
  \institution{Aalto University}
  \city{Espoo}
  \country{Finland}
}
\email{shuai.ma@aalto.fi}

\author{Junling Wang}
\affiliation{
  \department{}
  \institution{ETH Zürich}
  \city{Zürich}
  \country{Switzerland}
}
\email{junling.wang@ai.ethz.ch}

\author{Yuanhao Zhang}
\affiliation{
  \department{}
  \institution{The Hong Kong University of Science and Technology}
  \city{Hong Kong}
  \country{China}
}
\email{yzhangiy@connect.ust.hk}

\author{Xiaojuan Ma}
\affiliation{
  \department{}
  \institution{The Hong Kong University of Science and Technology}
  \city{Hong Kong}
  \country{China}
}
\email{mxj@cse.ust.hk}

\author{April Yi Wang}
\affiliation{
  \department{}
  \institution{ETH Zürich}
  \city{Zürich}
  \country{Switzerland}
}
\email{april.wang@inf.ethz.ch}

\renewcommand{\shortauthors}{Shuai Ma, et al.}

\begin{abstract}

Decomposition is a fundamental skill in algorithmic programming, requiring learners to break down complex problems into smaller, manageable parts. However, current self-study methods, such as browsing reference solutions or using LLM assistants, often provide excessive or generic assistance that misaligns with learners' decomposition strategies, hindering independent problem-solving and critical thinking. To address this, we introduce Decomposition Box (DBox), an interactive LLM-based system that scaffolds and adapts to learners' personalized construction of a step tree through a \emph{``learner-LLM co-decomposition''} approach, providing tailored support at an appropriate level. A within-subjects study (N=24) found that compared to the baseline, DBox significantly improved learning gains, cognitive engagement, and critical thinking. Learners also reported a stronger sense of achievement and found the assistance appropriate and helpful for learning. Additionally, we examined DBox's impact on cognitive load, identified usage patterns, and analyzed learners' strategies for managing system errors. We conclude with design implications for future AI-powered tools to better support algorithmic programming education.

\end{abstract}



\maketitle

\section{Introduction}

Algorithmic programming, which involves applying algorithms to solve real-world problems, is a challenging yet essential skill for computer science learners \cite{backhouse2011algorithmic, skiena1998algorithm}. Unlike introductory programming, the primary challenge in algorithmic programming lies not in algorithmic concepts, basic logic or syntax but in decomposing a complex problem to develop a holistic solution \cite{smetsers2017problem, backhouse2011algorithmic}. Students often struggle with formulating clear strategies, getting stuck at specific steps, or overlooking edge cases \cite{linn1985cognitive, piech2015autonomously}. Online knowledge communities (e.g., LeetCode) are commonly used for self-study, where learners can browse reference solutions provided by these platforms or other online resources (e.g., search engines or Q\&A platforms) when encountering difficulties. However, these resources are not tailored to individual problem-solving approaches and can hinder independent thinking and learning.

In addition to traditional resources, the rapid advancement of large language models (LLMs) has introduced AI tools that rival or even surpass human performance in certain programming tasks \cite{finnie2023my}. Recent studies in CS education and human-computer interaction (HCI) have explored the impact of LLM-based tools like ChatGPT, GitHub Copilot, and Codex on introductory programming (CS1) \cite{kazemitabaar2023studying, jin2024teach, finnie2022robots, roest2024next}. While LLMs are increasingly used for tasks such as code completion, translation, debugging, summarization, and explanation \cite{hou2024large, sarsa2022automatic}, they are not specifically designed for educational purposes, leading to issues like over-reliance and a lack of independent problem-solving \cite{kazemitabaar2023novices, sheese2024patterns}, as highlighted in our formative study observing programming learners solving algorithmic problems using various AI support tools. Moreover, most research has focused on the impact of LLM on novice learners in introductory programming \cite{kazemitabaar2023studying, jin2024teach, finnie2022robots, roest2024next}, with little attention given to how these tools can be tailored to support algorithmic programming learning. \ms{Therefore, we pose the following research question: \textbf{How can LLM-supported interfaces be designed to effectively facilitate pedagogically meaningful learning in algorithmic programming?}}

In this paper, we present the Decomposition Box (DBox), a tool that leverages LLMs to assist learners in decomposing algorithmic programming problems. DBox adopts a \emph{Learner-LLM Co-Decomposition} design, where learners lead the decomposition process while the LLM provides scaffolding only when needed. DBox supports two key stages of algorithmic programming: solution formation and implementation. During the solution formation stage, DBox introduces an interactive step tree that adapts to students' existing code or natural language thought processes. Students can express and update their ideas by directly writing code or iteratively constructing the step tree using interactive features, with DBox providing real-time feedback on the status of each node. In the implementation stage, DBox validates the alignment between learners' code and the step tree, identifying the status of each node in real time. A progressive hint system further supports learners by starting with thought-provoking questions for incorrect or missing steps and gradually offering more specific hints after repeated attempts. This design balances problem-solving progress with fostering independent thinking, with the hint system seamlessly integrated across both stages.

To evaluate DBox's effectiveness in supporting algorithmic programming learning, we conducted a within-subjects mixed-methods user study with 24 learners.
Our analysis highlighted that DBox significantly improved students' programming performance, perceived learning gains, confidence, algorithmic thinking, and self-efficacy. 
Students reported greater cognitive engagement, critical thinking, and a stronger sense of achievement compared to the baseline, where many felt they were ``cheating'' or not solving problems independently. 
DBox was seen as offering more appropriate assistance and benefiting learning. 
Since DBox relies on LLM for assessments, we also conducted a technical evaluation to assess the quality of LLM prompts used in the pipeline. 
This evaluation demonstrated that LLM are sufficiently accurate to support DBox's features, particularly in identifying incomplete and incorrect code approaches, though some challenges were noted in handling natural language descriptions.

Building on the key findings from our experiment, we provided several design implications for developing tools that promote algorithmic programming learning. We also discussed DBox's design philosophy, particularly the human-AI co-decomposition approach, and explored how to effectively leverage LLM to support programming learning, even with LLM's imperfection. In summary, our contributions are as follows:


\begin{itemize}
    \item A formative study that identified four challenges learners face in algorithmic programming learning with existing tools, leading to four design goals.
    \item The design of DBox, an AI-assisted algorithmic programming learning tool, features a scaffolded interactive step tree that facilitates learner-AI co-decomposition of problems, supporting both the ideation and implementation stages while fostering independent thinking and active learning.
    \item A technical evaluation of LLM' accuracy in assessing students' fine-grained thought processes, highlighting where LLMs excel and where they face challenges.
    \item A user study reported DBox's effectiveness on algorithmic programming learning, including its effects on learners' learning outcomes, perceptions, user experience, and usage patterns.
\end{itemize}

\section{Related Work}





\subsection{Scaffolding Programming Learning}

Scaffolding, rooted in constructivist theory, helps bridge the gap between learners' current abilities and desired skills by integrating new information into existing cognitive frameworks, rather than passively receiving it \cite{kim2011scaffolding, pea2018social, cole1978mind, wood1976role, tobias2009constructivist}.
Learners with limited knowledge particularly benefit from scaffolding to enhance understanding and avoid floundering \cite{hmelo2007scaffolding, schmidt2007problem}. 
In programming education, scaffolding helps structure the learning process, guiding learners through problem-solving via hints or correcting misconceptions \cite{kim2011scaffolding, sykes2010design}.
Traditional scaffolding often relies on human-provided support, but recent technological advancements have expanded its scope to include digital solutions \cite{delen2014effects, kim2011scaffolding}. For example, methods such as using flowcharts to brainstorm and organize solution ideas have been shown to improve algorithm design and programming skills \cite{smetsers2017problem}. Similarly, Cunningham et al. describe a multi-stage programming process that includes explicit planning, offering a structured approach to learning \cite{cunningham2021avoiding}. 
Generally, providing immediate assistance during code writing—such as detailed feedback on errors, suggestions for corrections, or next-step hints—illustrates key scaffolding strategies that effectively enhance learner understanding \cite{singh2013automated, sykes2010design, rivers2017data}. Other supportive methods include the use of worked examples \cite{wang2022exploring} and Parsons problems, which engage students in active problem-solving \cite{hou2022using}.

Adaptive scaffolding, which adjusts support based on real-time feedback from learners, is particularly effective. This approach tailors instruction to the learner’s evolving understanding and knowledge level, though it poses significant challenges for tools that require cognitive models to interpret learner input \cite{aleven2002effective, conati2000toward, vanlehn2011relative}.
Recent advancements in large language models (LLMs) have introduced opportunities for adaptive scaffolding by inferring learner's mental state from their inputs, enabling adaptive, learner-specific assistance. \emph{However, most LLM applications in scaffolding have been limited to generating code explanations and next-step hints \cite{macneil2023experiences, roest2024next}, with little focus on adaptive scaffolding tailored to individual needs during algorithmic programming}. Our research addresses this gap by using LLMs to enhance the Zone of Proximal Development (ZPD) \cite{chaiklin2003zone} in algorithmic programming through carefully crafted prompts. We capture students' thought processes in both code and verbal forms, forming a step-tree structure that reflects their current thinking. This model facilitates targeted assistance, assessing and intervening at various levels for each identified error or gap. By ensuring that scaffolding aligns with the learner's current knowledge state and providing only essential guidance, our approach promotes independent thinking and substantially improves learning outcomes.

\subsection{AI Coding Assistants and Application in Educational Contexts}

LLMs have demonstrated their capabilities in programming-related tasks, including delivering precise feedback on syntax errors \cite{phung2023generating} and enhancing programming workflows \cite{leinonen2023using}. Sarsa et al. evaluated OpenAI Codex’s potential for generating engaging programming exercises \cite{sarsa2022automatic}, while MacNeil et al. observed that student engagement with LLM-generated code explanations varied depending on the complexity and length of the content \cite{macneil2023experiences}. Additionally, Prather et al. highlighted the dual-edged nature of GitHub Copilot’s suggestions, cautioning against potential dependency issues and underscoring the importance of fostering meta-cognitive skills in novice programmers \cite{prather2023s}.

\ms{Recent work has also explored mapping natural language to code using LLMs, which aligns with aspects of our approach. For instance, Liu et al. \cite{liu2023wants} introduced grounded abstraction matching to help non-expert end-user programmers better understand LLM capabilities and refine their input for code generation. Their method iteratively converts natural language to code and back into grounded utterances to refine user input. While their focus is on improving language refinement, our work emphasizes scaffolding learners in decomposition and idea-building for solving algorithmic problems, placing less priority on natural language quality.}

As LLMs advance in code generation, their applications in programming education have garnered increasing attention, particularly for creating educational content, enhancing student engagement, and personalizing learning experiences \cite{kasneci2023chatgpt}. Recent studies have examined these opportunities, focusing on task completion \cite{denny2023conversing}, instructional content generation \cite{leinonen2023using}, and innovative methods for content creation \cite{denny2022robosourcing}. Notably, Finnie-Ansley et al. found that OpenAI Codex outperformed most students in coding tasks during CS1 and CS2 exams \cite{finnie2023my}. Similarly, Kazemitabaar et al. studied how AI code generators like Codex support novice learners, showing increased code-authoring performance without compromising manual code-modification skills \cite{kazemitabaar2023novices}.

While much of the existing work focuses on introductory programming, our research addresses the challenges of advanced algorithmic programming, such as those encountered in CS2 courses. One closely related piece of work is that of Jin et al. \cite{jin2024teach}, who designed a teachable agent based on LLMs, employing a learning-by-teaching approach to help students grasp algorithmic concepts. However, their focus is primarily on the mastery of algorithmic concepts, whereas our work addresses the difficulties students face in applying these concepts to solve practical problems. Another tool similar to our work is Code Tutor \cite{codetutor}, which helps students tackle programming problems by guiding their thinking without giving direct answers. However, these conversational LLM tools differ from DBox in two key ways. First, their chat-based format can lead to students getting lost in lengthy conversations, whereas DBox uses an interactive step tree to build a structured mental model. Second, these tools rely on students inputting code or questions disconnected from their existing work, while DBox integrates the step tree with the editor, providing better context understanding and reducing extra input, thereby improving learning efficiency.

\subsection{Computational Thinking and Problem Decomposition}

\ms{Computational Thinking (CT), as originally conceptualized by Wing \cite{wing2006computational}, includes essential cognitive processes such as decomposition, abstraction, generalization, algorithmic thinking, and debugging \cite{angeli2016k}. Among these, decomposition is widely recognized as a cornerstone of CT. Early studies, such as those by Roy Pea, emphasized how programming environments like LOGO foster planning and decomposition skills by encouraging learners to break problems into smaller, manageable subproblems \cite{pea1987logo, pea1984cognitive}.} McCracken et al. \cite{mccracken2001multi} further highlighted its significance through a five-step learning framework for CS1 courses: (1) abstracting the problem from its description, (2) generating subproblems, (3) transforming these into sub-solutions, (4) re-composing them into a complete program, and (5) evaluating and iterating.

The importance of decomposition is also reflected in various educational strategies. For example, Keen and Mammen implemented a term-long course project that initially provided explicit guidance on program decomposition, gradually reducing support as the course progressed \cite{keen2015program}. Similarly, Sooriamurthi designed an exercise for CS1 students that required segmenting a large programming task into smaller components, promoting abstraction and iterative development skills \cite{sooriamurthi2009introducing}. Pearce et al. adopted a guided inquiry-based learning method in CS1, explicitly teaching decomposition strategies and using a rubric to assess students’ skills. Their findings demonstrated that students who received explicit scaffolding showed greater proficiency in breaking down problems \cite{pearce2015improving}.

Despite the well-documented value of decomposition in cultivating CT and programming skills, few tools leverage LLMs for problem decomposition. \ms{One related work by Kazemitabaar et al. \cite{kazemitabaar2024improving} explored task decomposition in data analysis programming, helping users break tasks into substeps or subphases to better steer and validate LLM-generated assumptions and code. However, their approach depends on LLMs automatically generating decomposition solutions for users to review. In contrast, our work emphasizes scenarios where learners actively develop decomposition skills, with LLMs providing minimal scaffolding. We term this approach ``learner-LLM co-decomposition'', fundamentally distinct from the ``LLM-generate then user-verify'' paradigm.}

To mitigate the aforementioned gaps, this paper introduces DBox, a tool designed to help students break down complex problems into manageable steps using a structured step tree. Through learner-AI co-decomposition, learners actively build the step tree while receiving step-level feedback from the LLM, enhancing their computational thinking and problem-solving skills.

\section{Formative Study}
In addition to traditional programming support tools like LeetCode, search engines, and Q\&A sites like Stack Overflow, AI-assisted tools such as ChatGPT and GitHub Copilot have further enriched these resources. 
However, it remains unclear whether these tools effectively enhance algorithmic programming learning and whether challenges persist despite their availability.
To explore this, we conducted a formative study using contextual inquiry and interviews to understand learner's needs and obstacles. 
Our study focuses on students who have completed foundational computer science courses and are working to improve their algorithmic problem-solving skills. 

\subsection{Study Procedure}
We recruited 15 university students (5F, 10M, aged 18-29), including 10 undergraduates and 5 graduate students. 
Most (11) were computer science majors, with the rest from related fields such as data science and electrical engineering. 
All participants had prior experience with LeetCode or similar platforms, 14 had used ChatGPT for programming tasks, and eight had experience with GitHub Copilot or similar tools. 
Each session lasted 40 minutes with \$10 compensation.

During the study, participants used tools like LeetCode, search engines, ChatGPT, and GitHub Copilot to solve a randomly assigned algorithmic problem for 20 minutes, during which we observed their tool usage and conducted contextual inquiries as notable behaviors arose. Afterward, we reviewed their activities and conducted a semi-structured interview to discuss their tool choices, usage, assistance received, and perceptions of the tools, including any benefits or drawbacks they experienced.

\subsection{Data Analysis and Results}
We conducted inductive thematic analysis \cite{hsieh2005three} on interviews and contextual inquiry data. Recorded sessions were transcribed and manually reviewed and corrected by the first author. 
Two authors then independently coded the data and discussed to finalize themes and categorizations. 
The analysis revealed four key challenges students face with existing assistance tools during algorithmic programming learning.

\ms{
\subsubsection{\textbf{Challenge 1: Excessive Help Hindering Active Learning}}
Students expressed concerns that platforms like LeetCode and ChatGPT provide solutions that are too easily accessible, hindering their ability to engage in independent problem-solving. 
While learners typically need minimal guidance to overcome specific hurdles, these tools often present complete solutions prematurely, making the learning experience ``\emph{uninteresting}'' and ``\emph{unfulfilling}'' (P3). As P1 noted, ``\emph{I just wanted help with the step I'm stuck on, but the solution panel [in Leetcode] showed everything, making it hard to ignore what I didn't need.}'' Participants also found it difficult to use ChatGPT as a learning tool, as it often provided full solutions rather than encouraging active engagement. P5 explained, ``\emph{ChatGPT often directly showed the complete solution and code, even though I just asked GPT about the specific problem I was facing. Once I saw the solution, it was hard to ignore it. It felt like I was not really improving my programming skills.}'' Additionally, participants like P12 noted that making ChatGPT useful for active learning required significant effort to ``carefully craft prompt'', as a result they ``\emph{just copy the problem description directly into ChatGPT}''.
}

\ms{
\subsubsection{\textbf{Challenge 2: Misalignment Between Provided Solutions and Learners' Own Problem-Solving Approaches}}
Students often struggled when provided solutions failed to align with their intended approaches or existing code. 
While learners preferred developing their own strategies, the solutions offered by platforms like LeetCode, search engines, or ChatGPT were often prescriptive that mismatched their efforts. 
Even when algorithms matched, differences in specifics made integration challenging, especially with partial or incorrect code. 
Rather than starting over, learners wanted help tailored to their approach and context. As P5 explained, ``\emph{I prefer sticking to my own ideas. The provided solution takes a different approach, and I just need help tailored to my method. I don’t want to change my thinking to fit someone else's solution, even if it's better.}'' Similarly, P8 noted the frustration of aligning external solutions with her own work: ``\emph{I don't want to scrap my code and start anew, so I try to align the solution with my code line by line. It’s tedious to figure out which parts of the solution relate to what I’ve written, where my errors began, and what I missed.}''
}

\subsubsection{\textbf{Challenge 3: Lack of a Structured Problem-Solving Approach}}

Students highlighted the need for a structured method to break down complex algorithmic problems into manageable steps, such as initialization, handling edge cases, loops, and return values. Current tools often present solutions as large blocks of text or code without clear visual structure. As P13 noted when using LeetCode’s solution panel, ``\emph{The solution outlines six steps, but the content is very disorganized and the text is unclear, making it hard to follow a structured process.}'' Some participants preferred step-by-step interactions with ChatGPT, asking specific questions as they progressed. However, resolving issues often required lengthy, multiple rounds of conversations, making it difficult to stay organized. Learners frequently had to scroll back and forth to find points related to specific steps. As P15 (using ChatGPT) shared, ``\emph{I had a detailed conversation with GPT spanning several pages. Scrolling back to revisit earlier points was cumbersome, and despite getting answers, it was hard to organize them into a coherent thought process}.''

\subsubsection{\textbf{Challenge 4: Insufficient Fine-Grained Feedback on Learners' Coding Progress}}

LeetCode's ``run code'' button evaluates the entire solution and only provides feedback on the overall correctness. This all-or-nothing approach overlooks partial correctness and fails to highlight specific errors, making it difficult for students to track their progress or verify their understanding step by step. For example, P8 noted, ``\emph{I need to know if a step is correct before I can decide how to proceed. Without step-by-step feedback, I can only keep doing it until I complete all the steps.}'' Additionally, students wanted early validation of their thought process before committing to code. Ten participants expressed the need for a feature to confirm if their approach was on the right track. As P1 stated, ``\emph{The correctness of my initial thought process is crucial. I hesitate to implement code until I'm confident it's correct. There's no tool to validate my approach early on.}''

\subsection{Design Goals} \label{designgoal} Based on the challenges identified in our formative study, we established the following design goals for our tool:

\begin{itemize}
    \item \textbf{D1: Scaffolding for Active Learning and Independent Thinking}: To address Challenge 1, our system should provide scaffolding support. Instead of presenting complete solutions, it should offer tailored support that encourages active problem-solving and independent thinking.
    
    \item \textbf{D2: Personalization to Individual Problem-Solving Styles}: In response to Challenge 2, our system should adapt to each student's unique problem-solving approach by analyzing how they break down problems and offering personalized feedback that preserves and enhances learners' own strategies.

    \item \textbf{D3: Connection and Structured Solution Presentation}: To address Challenge 2\&3, our system should visually connect the system-generated guidance to students’ existing codes and solutions, promoting a structured problem-solving approach that aligns with their mental models.

    \item \textbf{D4: Fine-Grained Evaluation and Feedback}: Addressing Challenge 4, our system should evaluate the correctness of students' thought processes—whether conveyed through code, pseudocode, or natural language—and provide detailed, step-by-step feedback to support continuous improvement.
\end{itemize}

\section{Decomposition Box}

\subsection{Overview} \label{overview}

\begin{figure*}[htbp]
	\centering 
	\includegraphics[width=\linewidth]{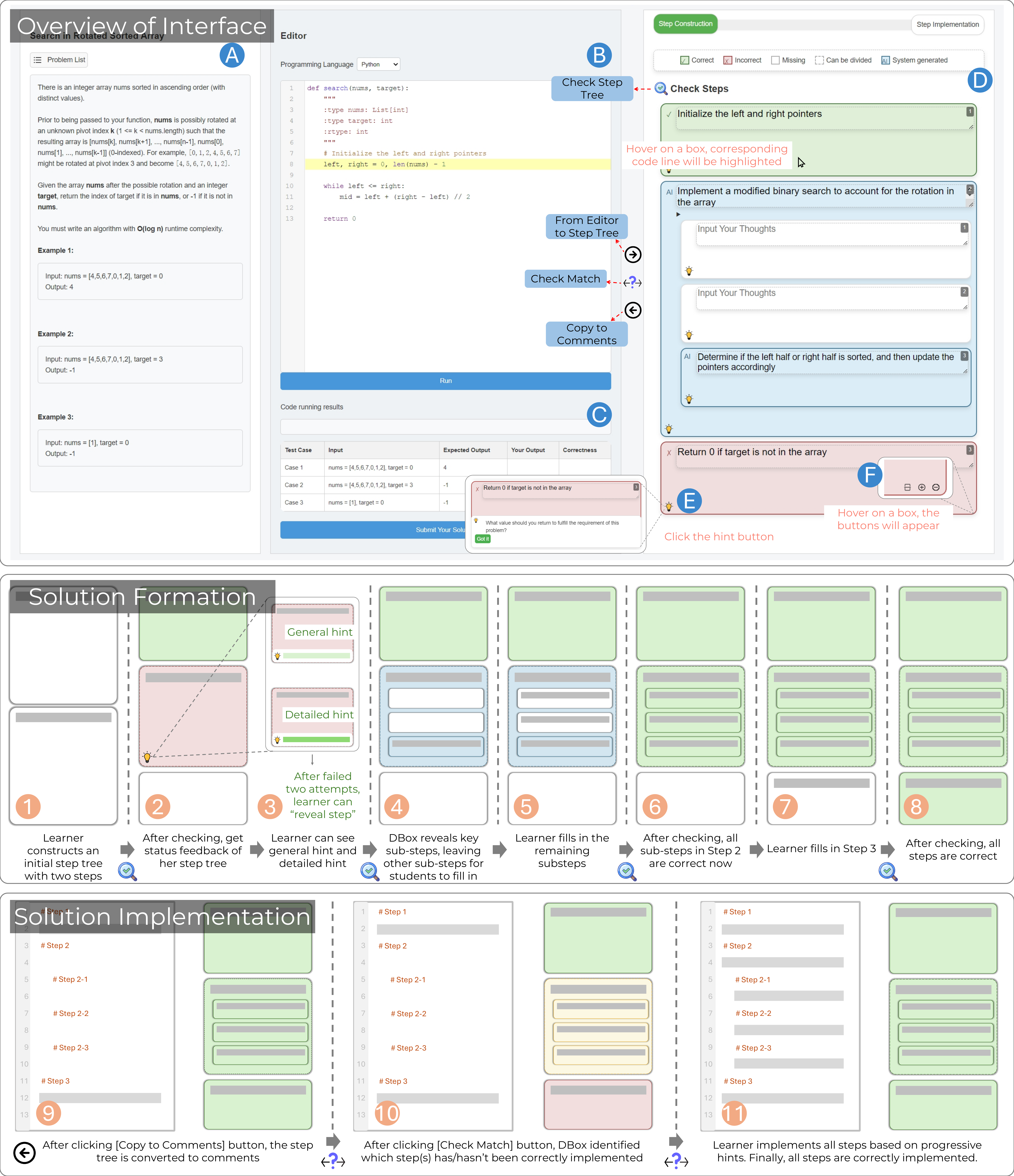}
	\caption{The interface of Decomposition Box. The top row displays the full interface in the solution formation stage (the solution implementation stage is similar, but with different status indicators). The middle row demonstrates a learner's solution formation stage showing basic DBox features. The bottom row illustrates a learner's solution implementation stage. An overview of the DBox interface and workflow is provided in Sec. \ref{overview}, and an illustrative example is described in Sec. \ref{illustrative}. To save space, the second row omits the problem description and editor area, and the third row excludes the problem description area.}
	\label{fig:interface}
        \Description{}
\end{figure*}

\begin{figure*}[htbp]
	\centering 
	\includegraphics[width=\linewidth]{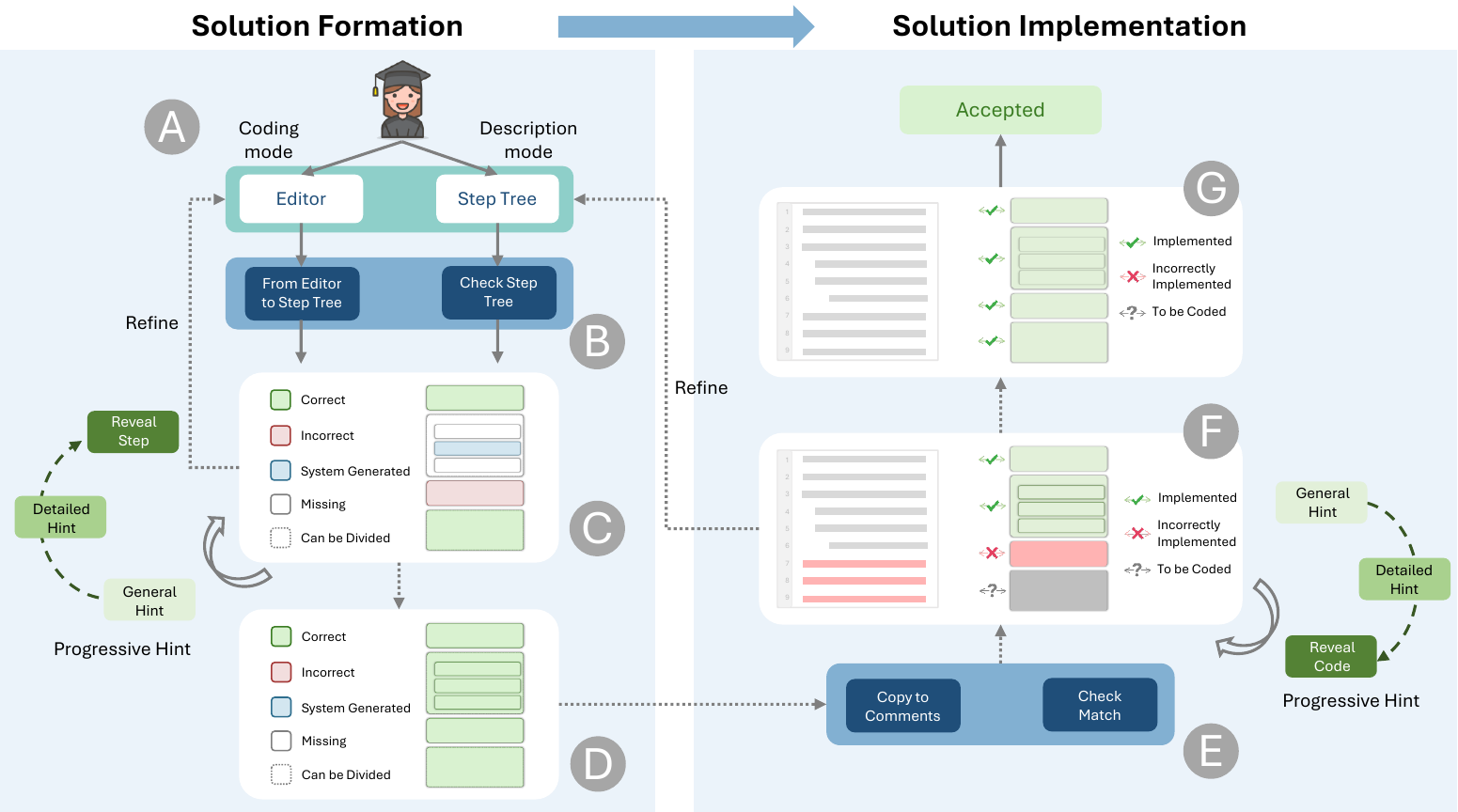}
	\caption{The DBox workflow supports learners through solution formation and implementation stages. During solution formation, (A) students can input ideas by either coding or using natural language to build a step tree. (B) By clicking ``From Editor to Step Tree'' or ``Check Step Tree'', (C) DBox renders the step tree and identifies node statuses (e.g., correct, incorrect, missing). Students can iteratively refine their code or step tree, receiving progressive hints, (D) until the step tree is fully correct. In the solution implementation stage, (E) students can convert the step tree into code comments or (F) check the alignment between their code and the step tree. Each node displays one of three statuses, and students can refine their work with ongoing hints until (G) all nodes are marked as ``implemented''.
 Finally, students can test if their code passes all test cases.}
	\label{fig:workflow}
        \Description{}
\end{figure*}

As shown in Figure \ref{fig:interface} (top row), DBox's interface has three main parts. The Problem Description (Figure \ref{fig:interface}.A)) and Solution Code Editor (Figure \ref{fig:interface}.B) are similar to the LeetCode platform.
The Interactive Step Tree Widget (Figure \ref{fig:interface}.D) enables users to refine their thought process and receive feedback via an interactive step tree. 
Three buttons—``From Editor to Step Tree'', ``Check Match'', and ``Copy to Comments''—connect the editor and step tree. 
The ``Check Step Tree'' button provides feedback on the step tree's status, categorized into five types (Figure \ref{fig:interface}.D). 
Clicking ``Hint'' offers progressive guidance based on learners' existing attempts (Figure \ref{fig:interface}.E). 
Hovering over steps shows buttons for editing the step tree (Figure \ref{fig:interface}.F).
Figure \ref{fig:workflow} shows two key stages in DBox's workflow:
\begin{itemize}
    \item \textbf{Solution Formation}: 
    The step tree starts as an empty box where students can freely add steps in either coding mode (directly writing code) or description mode (building a step tree using natural language). 
They can evaluate their progress with the ``From Editor to Step Tree'' or ``Check Step Tree'' buttons. The tree contains steps and substeps labeled as \emph{Correct}, \emph{Incorrect}, \emph{System Generated}, \emph{Missing}, or \emph{Can be Divided}. 
Layouts adjust dynamically based on the hierarchy. 
Steps that can be further divided are marked with dashed outlines, serving as a reminder, though students can decide whether further division is necessary. 
    \item \textbf{Solution Implementation}: Students can convert the step tree into comments with “Copy to Comments” or verify alignment by clicking “Check Match”. Nodes in the step tree are labeled as Implemented, Incorrectly Implemented, or To Be Coded. In this stage, DBox also offers progressive hints. When all nodes are implemented, students can test their solution by clicking “Run” button against the provided test cases.
\end{itemize}

\subsection{Target Users and A System Walkthrough} \label{illustrative}
\ms{DBox is designed for learners who understand basic algorithm concepts but struggle to apply them to solve practical problems. Using a scaffolding approach, DBox emphasizes independent thinking by offering only essential support. It assumes students are motivated, self-regulated, and actively engaging with the tool to improve their decomposition skills. If a student is less motivated or prefers a quicker solution, they may bypass DBox to search for answers online.}
Next, we present an example walkthrough (Figure \ref{fig:interface}) of such a self-regulated student Alice: 

Alice, a learner tackling the ``Search in Rotated Sorted Array'' problem, begins by organizing her thoughts in the solution formation stage. DBox offers two options: she can either start coding or build a step tree using natural language. She opts for the latter and adds two initial steps (Figure \ref{fig:interface}.1). To check her progress, Alice clicks ``Check Step Tree'' button. DBox flags Step 1 as correct, Step 2 as incorrect, and highlights a missing Step 3 (Figure \ref{fig:interface}.2). She clicks the hint button on Step 2, receiving general and detailed guidance, but after another failed attempt, DBox offers another option for revealing a substep (Figure \ref{fig:interface}.3). Alice clicks ``Reveal (Sub)Step'', uncovering a sub-step 2-3 while leaving sub-steps 2-1 and 2-2 for her to solve (Figure \ref{fig:interface}.4). Inspired by the hints, Alice figures out how to break down and fills in these sub-steps (Figure \ref{fig:interface}.5). After checking again, Step 2 is marked correct (Figure \ref{fig:interface}.6). Alice adds the missing Step 3 (Figure \ref{fig:interface}.7), and finally, after checking, all steps turn to correct (Figure \ref{fig:interface}.8).

Next, Alice moves to the solution implementation stage. She clicks ``Copy to Comments'', and DBox converts her step tree into code comments (Figure \ref{fig:interface}.9). As Alice writes her code, she uses the ``Check Match'' button to identify incorrectly implemented and unimplemented steps. Step 2 is identified as unimplemented and Step 3 is identified as incorrectly implemented (Figure \ref{fig:interface}.10). Following DBox's guidance, she revises the code, and after another check, all steps turn to be correctly implemented (Figure \ref{fig:interface}.11). Satisfied with her progress, Alice clicks ``Run'' and successfully passes all test cases, solving the problem.

Note that we have presented only a simple walkthrough here, whereas the steps in a student's actual problem-solving process are more complex and dynamic (as shown later in Sec. \ref{actual_use}). Next, we introduce the specific features aligned with the four design goals as described in Sec. \ref{designgoal}.

\subsection{Stage 1: Solution Formation}
\subsubsection{Two Input Modes (D1, D2, D3)}
DBox offers users the flexibility to develop their solutions through two distinct input modes: by writing code directly or by constructing a step tree using natural language descriptions, without needing to start with code. In the latter mode, users begin with a blank step tree and can click ``Add'' to insert nodes or ``Split'' to create sub-steps for more granular detail. Each node contains a text input field where users can articulate their thought process. Steps and sub-steps can be rearranged or deleted, allowing learners to iteratively and interactively refine and structure their mental model.

\subsubsection{Inferring Users' Thought Process from Existing Code (D1, D3)}
The ``From Editor to Step Tree'' function in DBox infers a learner’s intended solution and thought process based on their incomplete code. When activated, the system analyzes the code and problem, presenting the inferred steps as a tree on the right-hand side of the interface. Hovering over each node highlights the corresponding lines in the code editor, linking the inferred steps directly to the code. This feature assists users in diagnosing errors and identifying potential issues, especially when they are unsure how to proceed.

\subsubsection{Step Tree Node Status Evaluation with Preservation of Original Structure (D2, D4)}
When the learner clicks ``From Editor to Step Tree'' or ``Check Step Tree'', DBox evaluates each node, assigning one of five statuses:
(1) \textbf{Correct}: The step aligns with the learner's intended approach.
(2) \textbf{Incorrect}: Errors are identified in the step.
(3) \textbf{Missing}: A required step is absent.
(4) \textbf{Can Be Divided}: The step is complex and can be broken into sub-steps, indicated by dashed borders. Users decide whether to subdivide. This status can coexist with other statuses.
(5) \textbf{System Generated}: Step content is created by the system. This status is triggered only when the learner requests to reveal a (sub)step after repeated failures.
During the ``Check Step Tree'' process, DBox preserves the original step tree (both structure and contents), only adding blank nodes for missing steps, ensuring scaffolding while respecting the learner's thought process.

\subsubsection{Progressive Hints for Solution Formation (D1)}

DBox provides progressive hints to scaffold learners' problem-solving in three levels: (1) \textbf{General Hint} (Question-Based): Prompts learners' critical thinking without revealing solutions, e.g., ``Before converting the string to an array, what should you do first?'' (2) \textbf{Detailed Hint}: Offers more specific clues while requiring reasoning, e.g., ``Think about how you can traverse each character in the string.'' (3) \textbf{Reveal (Sub)Step} ((Sub)Step Recommendation): For repeated errors, the AI can suggest a substep within a larger step when users click the ``Reveal (Sub)Step'' button. This reveals one key substep while leaving the remaining steps for the learner to complete. Notably, students can choose not to trigger this hint. These progressive hints support problem-solving development while allowing learners to maintain independence and control.

Once the step tree is complete and all nodes are correct, learners proceed to the solution implementation stage.

\subsection{Stage 2: Solution Implementation}

\subsubsection{Converting the Step Tree into Comments (D3)} This feature converts each node of the step tree into code comments. When students click ``Copy to Comments'', the system intelligently inserts these comments into the appropriate sections of the code editor. This guides learners to implement their solutions within the corresponding parts of their code, ensuring a smooth transition from planning to coding while reinforcing their structured approach.

\subsubsection{Validating Code Implementation against the Step Tree (D3, D4)} The ``Check Match'' button evaluates the alignment between the code and the step tree. Steps are categorized and color-coded as: (1) \textbf{Implemented}, (2) \textbf{Incorrectly Implemented}, and (3) \textbf{To Be Coded}. Hovering over a step highlights the corresponding lines in the code, providing a direct mapping between the step tree and the code to help users efficiently debug their implementation.

\begin{figure*}[htbp]
	\centering 
	\includegraphics[width=\linewidth]{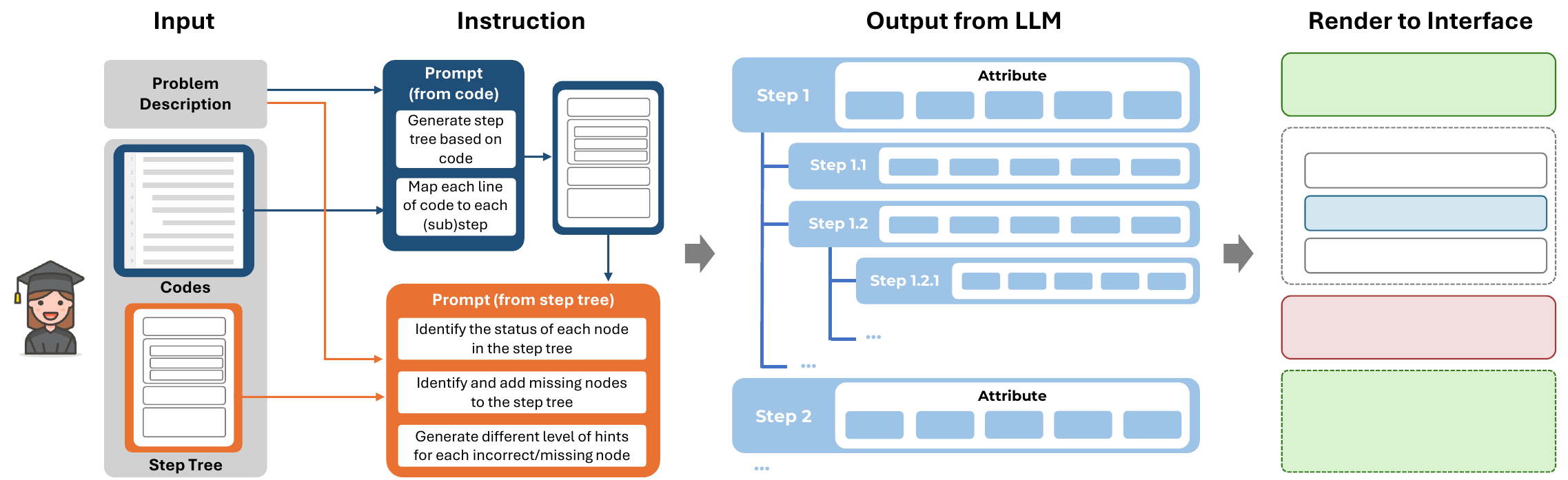}
	\caption{An illustration of DBox's data processing workflow highlights its core function—creating a step tree with node statuses from student inputs. The LLM processes learners' incomplete code or a step tree they’ve constructed. It outputs a structured JSON object containing steps, sub-steps (and sub-sub-steps, etc.), each with several attributes. Then the JSON object is rendered to the interface, preserving the original structure and only adding nodes for any missing steps. Each node keeps the student's original input, without directly revealing the correct solution. DBox encodes the status of each step with colors and provides progressive hints.}
 
	\label{fig:processing}
        \Description{}
\end{figure*}

\subsubsection{Progressive Hints for Solution Implementation (D1)}
For steps that are incorrectly implemented or yet to be coded, multi-level hints are available: (1) \textbf{General Hint}: Shows simple thought-provoking prompts/suggestions, e.g., ``How should you correctly iterate until the second last character?'' (2) \textbf{Detailed Hint} (Pseudocode): Provides simplified pseudocode to guide the user. (3) \textbf{Reveal Code} (Recommended Implementation): This option is activated only after two failed attempts. Clicking the ``Reveal Code'' button displays the recommended code implementation for the specific step.

\subsection{Backend Design}
DBox's backend is primarily powered by Large Language Models (the GPT-4o model specifically), with four distinct interactions corresponding to four buttons in the interface:
\begin{itemize}
    \item \textbf{From Editor to Step Tree}: This button sends the problem description and the user’s code to the LLM, which generates a step tree with nodes labeled as correct, incorrect, missing, or divisible.
    \item \textbf{Check Step Tree}: Clicking this button inputs the problem description and the user-constructed step tree into the LLM, which returns a labeled step tree with node statuses such as correct, incorrect, missing, or divisible.
    \item \textbf{Copy to Comments}: This button sends the problem description, current step tree, and user’s code to the LLM, generating a mapping of step tree nodes to corresponding lines of code.
    \item \textbf{Check Match}: Pressing this button sends the problem description, step tree, and user’s code to the LLM, which outputs a tree categorizing nodes as implemented, incorrectly implemented, or to be coded.
\end{itemize}


To illustrate, we present the data processing workflow for two core functions: ``From Editor to Step Tree'' and ``Check Step Tree''. Figure \ref{fig:processing} shows how DBox processes student inputs in two modes: coding mode and language description mode (step tree input). Prompts adapt based on the input type. When only code is provided (\textcolor{darkblue}{dark blue} lines in Figure \ref{fig:processing}), the system first calls \colorbox{darkblue}{\textcolor{white}{[prompt from code]}} to generate a step tree, mapping each code line to a corresponding (sub)step based on the code’s meaning. It then uses the generated step tree to invoke \colorbox{orange}{\textcolor{white}{[prompt from step tree]}} (\textcolor{orange}{orange} lines) to evaluate node statuses, add missing nodes, and generate multi-level hints for incorrect or missing nodes. If the input is a natural language step tree, the LLM directly calls \colorbox{orange}{\textcolor{white}{[prompt from step tree]}} while preserving the original structure. 

The LLM outputs a JSON containing steps, sub-steps, and subdivisions, each with attributes like student original input, status (e.g., correctness, completeness), LLM-validated content, and hints. The JSON is rendered conditionally, preserving the student’s original structure while highlighting missing or incorrect steps. Even if a student's input differs from what the LLM considers correct, the original content is preserved and marked with a status color. Hints are provided and triggered progressively, offering ``just the right'' level of guidance.

\ms{
\subsection{Implementation}
For the front-end, we used native HTML, JavaScript, and jQuery. On the back-end, we deployed the application with the Flask\footnote{https://flask.palletsprojects.com/en/stable/} framework on our university's server. The code editor utilized CodeMirror\footnote{https://codemirror.net/} integrated with pyodide.js\footnote{https://pyodide.org/en/stable/} for running Python code. We employed OpenAI's GPT-4o model with a temperature of 0.8 to maintain flexibility during scaffolding. To align better with the front-end's step tree, we set the response format by setting the parameter response\_format={``type'': ``json\_object''}, restricting the LLM's output. Prompts designed with the chain-of-thought (CoT) technique \cite{wei2022chain} are detailed in the supplementary materials.
}

\section{Technical Evaluation}
\label{techeval}

In DBox, the key feature is that learners construct a step tree using either code or natural language descriptions, while the LLM evaluates each step and provides necessary feedback. To assess whether effective prompt engineering enables the GPT-4o model (hereafter referred to as GPT or LLM) to accurately determine node statuses (i.e., correct, incorrect, or missing), we conducted a preliminary technical evaluation. Detailed prompts used are provided in the supplementary material.

\subsection{Dataset Creation} 
We created a dataset of learners' authentic thought errors to evaluate LLMs' ability to recognize the status of the thought process.
Based on GPT-4's performance on coding tasks~\cite{finnie2023my, savelka2023thrilled}, we selected 25 easy-level LeetCode problems covering various algorithms and data structures problems (e.g., dynamic programming, sorting, greedy algorithms).

To capture natural variations, we recruited five computer science students from a local university to manually create the step trees for five randomly selected problems (25 in total). 
Using correct code samples as references, annotators constructed a step tree based on their solution, including steps, substeps, and sub-substeps. 
They described each node in their own words and linked it to the relevant code.
After collecting the annotated step trees, we manually created various types of errors to simulate common student misconceptions in coding \cite{qian2017students}.
An algorithmic programming expert created seven error types for each problem (e.g., missing steps, incorrect step order, logical errors, and syntax errors), which were reviewed by another expert, resulting in 175 error-laden step trees (25 problems x 7 error types).

\subsection{Analysis Approach} We divide the steps into two parts based on the expert annotations: the correct part (1) or the incorrect/missing part (0). To calculate the performance of GPT, we use a very strict evaluation method: if all steps in the correct part are determined to be correct, the prediction of this part is marked as 1; otherwise, the prediction is 0. Similarly, if all steps in the incorrect/missing part are determined to be incorrect/missing, the prediction of this part is 0; otherwise, the prediction is 1. This approach enabled calculation of accuracy, F1 score, precision, recall, specificity, false positive rate, and false negative rate.
We removed status fields from 175 error-containing step trees and input them into GPT for prediction. The results were compared against expert-annotated ground truth. For incorrect predictions, two authors independently coded GPT's outputs to identify error themes and causes, later consolidated through discussion.

\begin{table*}[hbpt]
	\centering
	\caption{The technical evaluation of GPT-4o assesses its ability to identify the status of learners' steps. \textbf{Precision} refers to the proportion of steps correctly predicted as correct by GPT. \textbf{TPR} (True Positive Rate) measures the proportion of truly correct steps that GPT identifies correctly. \textbf{TNR} (True Negative Rate) reflects the proportion of truly incorrect/missing steps that GPT correctly predicts. \textbf{FPR} (False Positive Rate) indicates the proportion of incorrect/missing steps that GPT incorrectly predicts as correct. \textbf{FNR} (False Negative Rate) represents the proportion of correct steps that GPT incorrectly predicts as incorrect/missing.}
	\label{tab:technicalresult}%
	\begin{small}
	\begin{tabular}{c | c c c c c c c}
	    \hline
	    \textbf{Error Type}&Accuracy&F1&Precision&TPR/Recall&TNR/Specificity&FPR&FNR\\
	    \hline
 \hline
\multicolumn{8}{l}{\textbf{Identify step/substep status from learners' natural language-based step descriptions}}\\
\hline    
	    \textbf{Sequence Changed}&0.70&0.72&0.68&0.76&0.64&0.36&0.24\\
	
 \rowcolor{gray!15}\textbf{Logical Error}&0.88&0.87&0.91&0.84&0.92&0.08&0.16\\
	    \textbf{Missing}&0.86&0.86&0.85&0.88&0.84&0.16&0.12 \\

 \hline
\multicolumn{8}{l}{\textbf{Identify step/substep status from learners' codes}}\\
\hline
    \rowcolor{gray!15}\textbf{Sequence Changed}&1.00&1.00&1.00&1.00&1.00&0.00&0.00\\

	\textbf{Logical Error}&0.98&0.98&0.96&1.00&0.96&0.04&0.00\\

    \rowcolor{gray!15}\textbf{Missing}&0.92&0.92&0.92&0.92&0.92&0.08&0.08\\

    \textbf{Syntax Error}&0.90&0.90&0.88&0.92&0.88&0.12&0.08\\
	    \hline
	\end{tabular}%
	\end{small}
\end{table*}

\subsection{Results}
As shown in Table \ref{tab:technicalresult}, \textbf{GPT more accurately identifies students' thought processes when expressed through code rather than natural language}. This difference may stem from GPT's extensive code-based training data \cite{liu2024your} and specialized code-handling mechanisms \cite{achiam2023gpt}, while natural language descriptions often include imprecise or non-standard terminology, leading to ambiguities \cite{liu2023wants}.

\textbf{GPT sometimes identifies incorrect steps as correct (false positives) or correct steps as incorrect (false negatives)}. For \emph{sequence change errors}, GPT’s accuracy drops to 70\% with an F1 score of 72\% from natural language descriptions, with a False Positive Rate (FPR) of 36\% and a False Negative Rate (FNR) of 24\%. In contrast, code-based evaluations achieve 100\% accuracy and F1 scores. For \emph{logical errors} from natural language, GPT’s accuracy is 88\% (F1 score 87\%, FPR 8\%, FNR 16\%), compared to 98\% accuracy and F1 scores from code-based evaluations (FPR 4\%, FNR 0\%). \emph{Missing step error} evaluations from natural language yield 86\% accuracy (FPR 16\%, FNR 12\%), improving to 92\% from code inputs (FPR/FNR 8\%). \emph{Syntax error} identification remains steady at 90\% accuracy and F1 score.

Additionally, we find that \textbf{GPT occasionally alters the structure and content of the step tree}, despite instructions to only add missing steps. It sometimes modifies how steps are segmented or misinterprets the student's original input. Another key finding is that \textbf{GPT sometimes incorrectly judges non-standard approaches as wrong}. In five out of 25 tasks, GPT mistakenly flagged correct solutions as incorrect simply because they deviated from the common approaches in its training data. For example, it marked a sorting-based solution as incorrect, even though it was correct, albeit not the most optimal approach.

\ms{In summary, GPT demonstrates strong capabilities in processing code-based inputs but faces challenges with natural language, particularly in detecting sequence changes. Our analysis of participants' logs from the user study revealed that most errors encountered were logical errors or missing steps, while errors involving sequence changes were relatively uncommon. This suggests that GPT is generally effective in evaluating students' thought processes during algorithmic programming learning. We recommend that researchers considering GPT for supporting learners in natural language programming carefully evaluate its limitations and conduct technical assessments to determine its suitability for their specific scenarios. We hope this technical evaluation serves as a valuable reference for similar future research.}

\section{User Study}

We conducted a user study to evaluate the effects of DBox, focusing on three questions:
\begin{itemize}
    \item \textbf{Q1}: How does DBox support algorithmic programming learning?
    \item \textbf{Q2}: How does DBox affect learners' perceptions and user experience?
    \item \textbf{Q3}: How do learners interact with DBox and perceive the usefulness of different features?
\end{itemize}


\subsection{Conditions}
We conducted a within-subjects design to control for individual differences in programming abilities. Participants experienced two conditions in a randomly assigned order:

\begin{itemize}
    \item \textbf{DBox}: Participants solved problems using the proposed DBox.
    \item \textbf{Baseline}: Participants freely used any available tools (e.g., ChatGPT, Copilot, search engines, LeetCode, QA platforms) to reflect their real-world learning habits, with no restrictions on tool usage or combinations.
\end{itemize}

\subsection{Task and Materials}
In this experiment, participants solve problems from two distinct algorithm types. Each type includes a learning problem, where participants use DBox or baseline tools, and a test problem, solved independently without assistance.

We selected problems from the LeetCode problem bank based on several criteria: First, all problems were of medium difficulty, with an acceptance rate between 40\% and 50\% to ensure sufficient challenge. Second, GPT performs well on these problems. Third, the two algorithm types are distinctly different to avoid learning effects. Finally, the learning and test problems within each algorithm type require similar programming skills to avoid unfair comparisons due to differences in additional coding skills needed for each problem. Based on these criteria, we chose two algorithm types: Greedy and Binary Search. For Greedy, we selected ``Jump Game''\footnote{https://leetcode.com/problems/jump-game/description/} and ``Jump Game II''\footnote{https://leetcode.com/problems/jump-game-ii/description/}; for Binary Search, we selected ``Search in Rotated Sorted Array''\footnote{https://leetcode.com/problems/search-in-rotated-sorted-array/description/} and ``Search in Rotated Sorted Array II''\footnote{https://leetcode.com/problems/search-in-rotated-sorted-array-ii/description/}.

\ms{To help participants become familiar with or recall the algorithms used in the study, we provide them with lecture materials prior to the start of the study. The lecture materials for the two types of algorithms were sourced from GeeksforGeeks\footnote{Greedy: https://www.geeksforgeeks.org/introduction-to-greedy-algorithm-data-structures-and-algorithm-tutorials/}\footnote{Binary Search: https://www.geeksforgeeks.org/binary-search/}. These materials include an introduction to each algorithm, illustrated figures, and practical examples.}

\begin{figure*}[htbp]
	\centering 
	\includegraphics[width=0.9\linewidth]{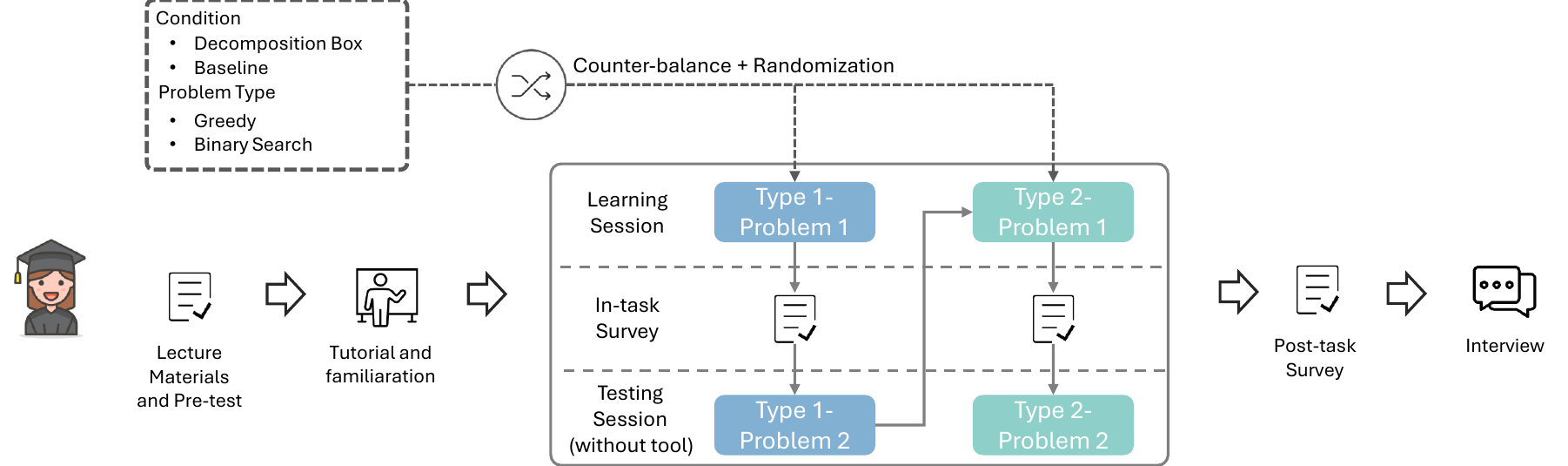}
	\caption{The procedure of our user study. \ms{To avoid learning effects, we used a counterbalanced design with four combinations: (1) DBox-Type1 → Baseline-Type2, (2) Baseline-Type1 → DBox-Type2, (3) DBox-Type2 → Baseline-Type1, and (4) Baseline-Type2 → DBox-Type1. For each combination, six participants were randomly assigned.}}
	\label{procedure}
        \Description{}
\end{figure*}

\subsection{Procedure}
\ms{As shown in Figure \ref{procedure}, obtaining participants' consent, we begin by explaining the study's objectives and procedure. We then familiarize participants with the algorithms they will be practicing using the provided lecture materials. Note that the lecture material was designed to help participants recap the key concepts of these two algorithms. Although the two problems are labeled as Binary Search and Greedy on the LeetCode platform, participants were not restricted to using these two specific approaches to solve the problems}. Afterwards, we administer a pre-test problem to assess their expertise. We also have participants rate their confidence in solving the problem on a 7-point Likert scale following \cite{jin2024teach}. \ms{After participants completed the task or indicated they could not proceed, the two authors (as experiment operators) assessed their solutions against pre-verified answers (with multiple solutions). Participants who solved the problem correctly were excluded, as it suggested higher expertise in the tested problem type. Following \cite{jin2024teach}, we also excluded participants who rated their confidence at 6 or above, as high self-confidence likely indicates less need for additional support. While self-reported confidence may not always align with actual ability, this criterion helps focus the study on the intended user group for our tool. We then provided participants with an interactive tutorial to familiarize them with DBox. The tutorial guides them through each view, button, and functionality of the tool. After the tutorial, they can explore the tool by solving an exercise problem, different from the two problem types in the main study.

Next, we assigned experimental conditions and problem types to participants with a counterbalanced design.} Within each problem type, one problem is randomly assigned in the learning session and the other in the testing session. In the learning session, participants use either DBox or baseline tools, during which they must successfully pass all test cases to proceed. They then complete an in-task survey about their perceptions and experience with the tool they just used. In the test session, participants solve problems without any tool assistance. After completing both problem types, they fill out a post-task survey, followed by a semi-structured interview.

\subsection{Participants}

We conducted a power analysis using G*Power \cite{faul2009statistical} for our two-condition within-subjects design. Assuming an effect size of $f = 0.6$ (moderate), $\alpha = 0.05$, and power of 0.8, we estimated a required sample size of 24 participants.

After IRB approval, we recruited 24 participants via emails and social media at local universities (10 female, 14 male, average age 23.5, SD = 1.7). The group included 16 undergraduates and eight graduate students, with majors in computer science (17), data science (3), electrical engineering (3), and mathematics (1). Most participants (18) coded weekly, six coded monthly, and 23 had used platforms like LeetCode. The 90-minute study compensated participants with 20 USD, equating to 13 USD/hour.

\subsection{Measurements}
Our measurements are summarized in Table \ref{tab:measurement}. To address \textbf{Q1} (effects on learning outcomes), we assessed correctness in the testing session \cite{kazemitabaar2023studying}, perceived learning gain \cite{zhou2021does}, confidence in solving similar problems \cite{hendriana2018role}, improvement in algorithmic thinking \cite{yaugci2019valid}, and self-efficacy \cite{tsai2019improving, yildiz2018digital}. For \textbf{Q2} (effects on perceptions and user experience), we measured cognitive engagement \cite{pitterson2016measuring}, critical thinking \cite{kamin2001measuring}, sense of achievement \cite{wiedenbeck2004factors}, feeling of cheating \cite{kazemitabaar2023studying}, perceived help appropriateness, usefulness \cite{davis1989perceived}, mental demand, effort, frustration \cite{hart2006nasa}, ease of use, satisfaction \cite{bangor2008empirical}, and future use intention \cite{holden2010technology}. For \textbf{Q3} (usage patterns and perceptions of DBox), we analyzed usage logs (e.g., clicks, edits, help-seeking), post-task feature ratings, and conducted semi-structured interviews to delve into participants' underlying reasons behind their perceptions, usage patterns, and reactions to AI errors. All questionnaires used a 7-point Likert scale.

\renewcommand{\arraystretch}{1.5}
\begin{table*}[htp]  

\centering  
\fontsize{8}{8}\selectfont  

\caption{Measurements used in our user study. For the questionnaire items (within the quotation marks), a 7-point Likert scale was used, with 1 indicating ``Strongly disagree/Very low'' and 7 indicating ``Strongly agree/Very high''.}\label{tab:measurement}

\begin{tabular}{m{0.5cm}<{\centering}m{3cm}<{\centering}m{10.5cm}}
\toprule
\textbf{}&\textbf{Metrics}&\textbf{Detailed Meaning and Questions}\\ \hline\hline

\multirow{6}*{\shortstack{\textbf{Q1}}}&Correctness Score & The correctness of learners' test task solutions was evaluated using a consistent rubric from \cite{kazemitabaar2023studying}. Two authors independently graded submissions, deducting 25\% for each major issue or missing component, yielding scores of 0\%, 25\%, 50\%, 75\%, or 100\%. They agreed on 87.5\% of submissions, resolving disagreements through discussion for the rest.\\
\cline{2-3}
&\cellcolor{gray!15}Perceived Learning Gain &\cellcolor{gray!15} "I have learned how to solve this type of problem." \\ 
\cline{2-3}
&Confidence in Solving Similar Problems & "After solving this problem with the tool's help, I feel confident in tackling similar problems." \\ 
\cline{2-3}
&\cellcolor{gray!15}Perceived Algorithmic Thinking Improvement & \cellcolor{gray!15}"This tool improved my ability to break down complex problems into smaller, manageable parts." \\ 
\cline{2-3}
&Self-Efficacy & "I have mastered the problem-solving skills necessary for this type of problem." \\ 
\hline

\multirow{12}*{\shortstack{\textbf{Q2}}}&\cellcolor{gray!15}Cognitive Engagement &\cellcolor{gray!15} "I was cognitively engaged in the programming exercises." \\
\cline{2-3}
&Critical Thinking & "The learning process challenged me to think critically." \\
\cline{2-3}
&\cellcolor{gray!15}Sense of Achievement &\cellcolor{gray!15} "I feel a sense of accomplishment/achievement when I complete the programming task." \\
\cline{2-3}
&Sense of Cheating & "Using this tool feels like cheating." \\
\cline{2-3}
&\cellcolor{gray!15}Perceived Appropriateness of Help &\cellcolor{gray!15} "I felt I received the right amount of help when needed—neither too much nor too little." \\
\cline{2-3}
&Perceived Usefulness & "This tool is useful for learning how to solve specific problems." \\
\cline{2-3}
&\cellcolor{gray!15}Mental Demand &\cellcolor{gray!15} "How mentally demanding was the task?" \\
\cline{2-3}
&Effort & "How hard did you have to work to achieve your level of performance?" \\
\cline{2-3}
&\cellcolor{gray!15}Frustration &\cellcolor{gray!15} "How insecure, discouraged, irritated, stressed, and annoyed were you?" \\
\cline{2-3}
&Ease of Use & "I find this tool easy to use for learning algorithms." \\
\cline{2-3}
&\cellcolor{gray!15}Satisfaction &\cellcolor{gray!15} "I am satisfied with the overall learning experience using this tool." \\
\cline{2-3}
&Future Use & "I would like to use this tool in my future programming learning." \\
\hline

\multirow{6}*{\shortstack{\textbf{Q3}}}&\cellcolor{gray!15}Button Clicking &\cellcolor{gray!15} (with timestamp) From Editor to Step Tree, Check Step Tree, Check Match, From Step Tree to Comments, Run Code\\
\cline{2-3}
&Editing & (with timestamp) Code edit, step tree edit\\
\cline{2-3}
&\cellcolor{gray!15}Help-Seeking &\cellcolor{gray!15} (with timestamp) see general hint, see detailed hint, and reveal step/code\\
\cline{2-3}
&Usefulness Rating &Participants' ratings on the usefulness of various features in DBox using a 7-point Likert scale\\
\cline{2-3}
&\cellcolor{gray!15}Interviews &\cellcolor{gray!15} Participants' detailed reasons for their perceptions, reactions to AI errors, and self-reported usage patterns, etc.\\

\bottomrule
\end{tabular}
\end{table*}

\subsection{Data Analysis}
\ms{To eliminate the unfair comparison caused by the learning effect of participants using both tools to solve the same type of problem, we selected two distinct problem types. We implemented a randomization procedure to ensure that each participant used either DBox or the baseline tool in a random order, with a randomly assigned problem type for each tool. As a result, each participant used only one tool to solve one problem.

For the quantitative analysis, we employed a linear mixed effects model to analyze the data. The dependent variables (DVs) were our outcome measures (e.g., scores or questionnaire ratings). First, we analyzed the main effect of the two different tools (the coefficient and p-value were reported based on this analysis). Then, we examined the interaction effects between the learning tool and problem type (\emph{Tool*Problem Type}), as well as the interaction effect between the learning tool and the order of tool usage (\emph{Tool*Order}). The fixed effects in the models included the learning tool, problem type, and the order of tool usage, while the random effect accounted for individual differences between participants.

}

For the qualitative analysis of our semi-structured interview data, grounded in the designed questions, we conducted a thematic analysis \cite{hsieh2005three}. Two authors independently coded the data, developed a codebook, and reached a consensus through discussion. In the results, we present key themes supported by representative participant quotes.


\section{Results}
In this section, we examine how DBox supports learners in algorithmic programming and how they interact with the tool. \ms{We compared DBox with the baseline tool and analyzed the interaction effects between tool and problem type, as well as the ordering effect (e.g., DBox first or Baseline first). Overall, we didn't find significant interaction or ordering effects for most metrics. Therefore, we report only the differences between the two tools unless notable interactions or ordering effects were observed, which are analyzed in detail.}

\subsection{How does Decomposition Box help with algorithmic programming learning?}

\begin{figure*}[htbp]
	\centering 
	\includegraphics[width=0.95\linewidth]{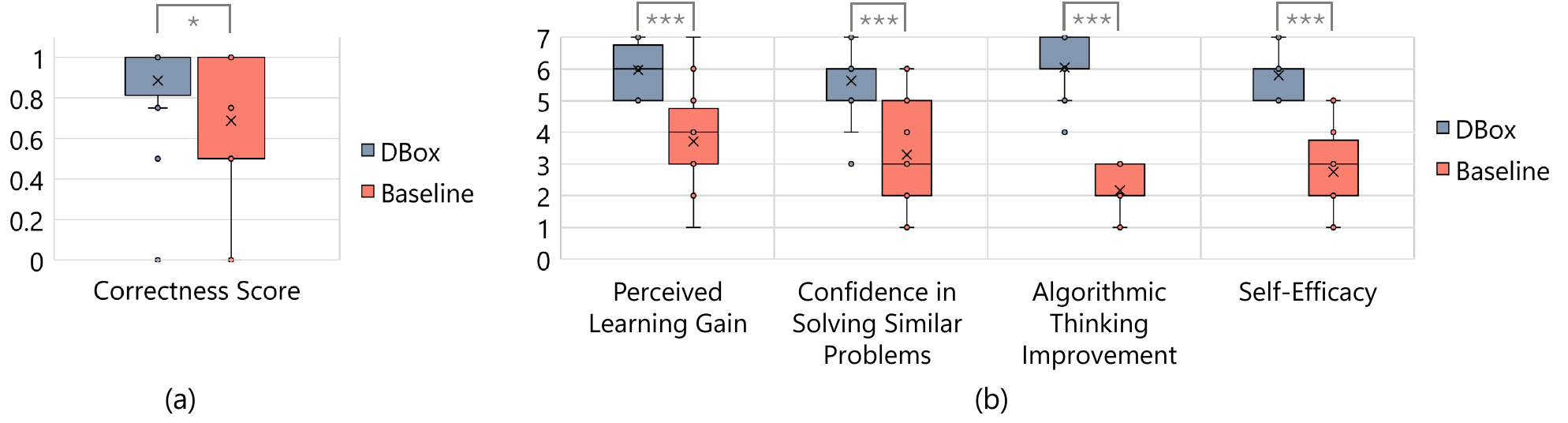}
	\caption{Effects on participants' learning outcomes: (a) Participants' correctness scores during the testing session, where they solved the problem independently. (b) Participants' self-reported metrics on their learning outcomes.}
	\label{RQ1}
        \Description{}
\end{figure*}

We first compared the correctness scores of participants' test session submissions under both DBox and baseline conditions. As shown in Figure \ref{RQ1} (a), participants using DBox achieved significantly higher correctness scores than those in the baseline condition ($Coef.$=0.198, $p$<0.05), suggesting that practicing with DBox better prepared learners to transfer their skills to similar algorithmic challenges.

This finding aligns with participants' subjective perceptions. Figure \ref{RQ1} (b) shows that learners in the DBox condition reported significantly higher perceived learning gains ($Coef.$=2.250, $p$<0.001), higher confidence in solving similar problems ($Coef.$=2.333, $p$<0.001), more improvements in algorithmic thinking ($Coef.$=3.875, $p$<0.001), and greater self-efficacy ($Coef.$=3.042, $p$<0.001) compared to the baseline condition.

Interview analysis further highlighted that participants felt solving tasks independently during the learning session enhanced their perceived learning gains. Overcoming challenges on their own also boosted their confidence. In contrast, baseline participants felt their algorithmic thinking was underdeveloped due to easy access to complete solutions (e.g., via search, ChatGPT, or Copilot), leading to lower perceived learning and confidence. As P5 noted, ``\emph{Even though I couldn’t write the full solution, the tool [DBox] encouraged me to break down the problem. I started with what I knew, and the tool guided me through the rest. Decomposing the problem helped me structure my approach, and as I saw the step tree fill in correctly, I felt my algorithmic thinking improve, and I gained confidence in solving the problem.}''

\subsection{How does Decomposition Box affect learners’ perceptions and user experience?}

\begin{figure*}[htbp]
	\centering 
	\includegraphics[width=0.95\linewidth]{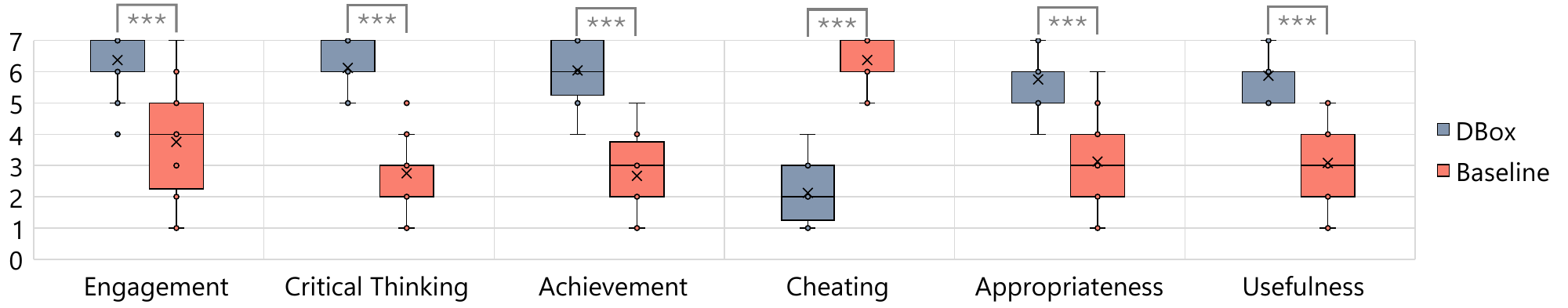}
	\caption{Participants' perceptions of the two conditions in their learning process.}
	\label{perception}
        \Description{}
\end{figure*}

\begin{figure*}[htbp]
	\centering 
	\includegraphics[width=0.95\linewidth]{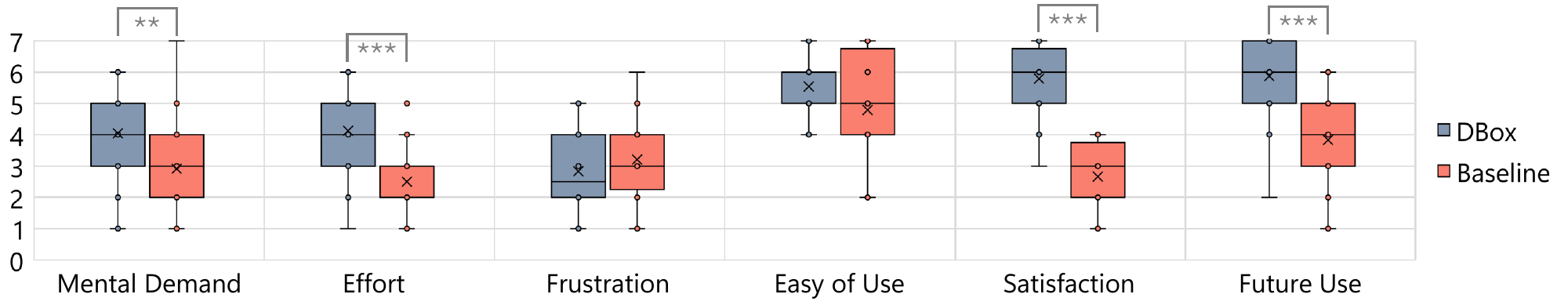}
	\caption{Participants' cognitive load and user experience.}
	\label{UX}
        \Description{}
\end{figure*}

\subsubsection{Effects on Learners' Perceptions}

As shown in Figure \ref{perception}, participants in the DBox condition reported significantly higher cognitive engagement ($Coef.$=2.625, $p$<0.001), greater critical thinking ($Coef.$=3.375, $p$<0.001), and a stronger sense of achievement ($Coef.$=3.375, $p$<0.001) compared to the baseline condition. Conversely, those in the baseline condition felt their problem-solving process resembled more ``cheating'' ($Coef.$=-4.250, $p$<0.001). DBox was also rated as providing more appropriate assistance ($Coef.$=2.625, $p$<0.001) and being more useful for programming learning ($Coef.$=2.792, $p$<0.001). Interviews supported these results, with 19 participants noting that DBox allowed them to independently develop their thought process while offering just enough feedback to guide them. This stimulated high engagement in problem-solving, as they critically analyzed both the overall solution and each step. As P1 noted, ``\emph{When I hit a block, unlike other tools (referring to the baseline), DBox didn’t give me the answer outright, which forced me to think through the problem myself. Even with help, I still had to do most of the thinking.}''

In contrast, baseline participants using tools like ChatGPT, Copilot, or LeetCode often \textbf{bypassed independent thinking, focusing on comparing or copying provided answers}. Twelve out of 24 compared answers while coding, and eight simply copied solutions, leading to a lack of achievement and a sense of ``cheating''. As P16 (using ChatGPT) admitted, ``\emph{I tried to convert its provided code into my own, but I didn’t feel like it was truly my solution; there was no sense of achievement.}'' Besides, \textbf{excessive help in the baseline tools led participants to feel it was unhelpful for learning}. For example, P23 (who used LeetCode’s built-in solution) shared, ``\emph{I was stuck on a small part, but the solution showed the entire answer immediately. I memorized it, but later, writing the code felt more like repetition than actual learning.}'' P15 (using ChatGPT) added, ``\emph{ChatGPT explained the problem and gave the full code. While its solution seemed right, I realized I was just judging its correctness rather than improving my programming skills.}''

\ms{Moreover, we found an interaction effect between tool and problem type ($Coef.$=1.250, $p$<0.05) in perceived usefulness. Post-hoc analysis showed that DBox outperformed the baseline in both Binary Search ($t=6.159$, $p<0.001$) and Greedy problems ($t=9.273$, $p<0.001$). With DBox, there was no significant difference in perceived usefulness between the two problems ($t=0.294$, $p=0.771$). However, with the baseline, perceived usefulness was lower for the Greedy problem compared to Binary Search ($t=-2.755$, $p<0.05$). We found no significant interaction effects on correctness scores, with participants showing similar performance across the two problems using either DBox or the baseline. This suggests that the lower perceived usefulness of the baseline for the Greedy problem was not due to the problem being inherently more difficult. A likely explanation is that the Greedy problem requires higher planning and decomposition skills (i.e., breaking a complex problem into subproblems solvable by a greedy algorithm) and the baseline did not provide scaffolding to support this, leading to lower perceived usefulness.
}

\begin{figure*}[htbp]
	\centering 
	\includegraphics[width=0.96\linewidth]{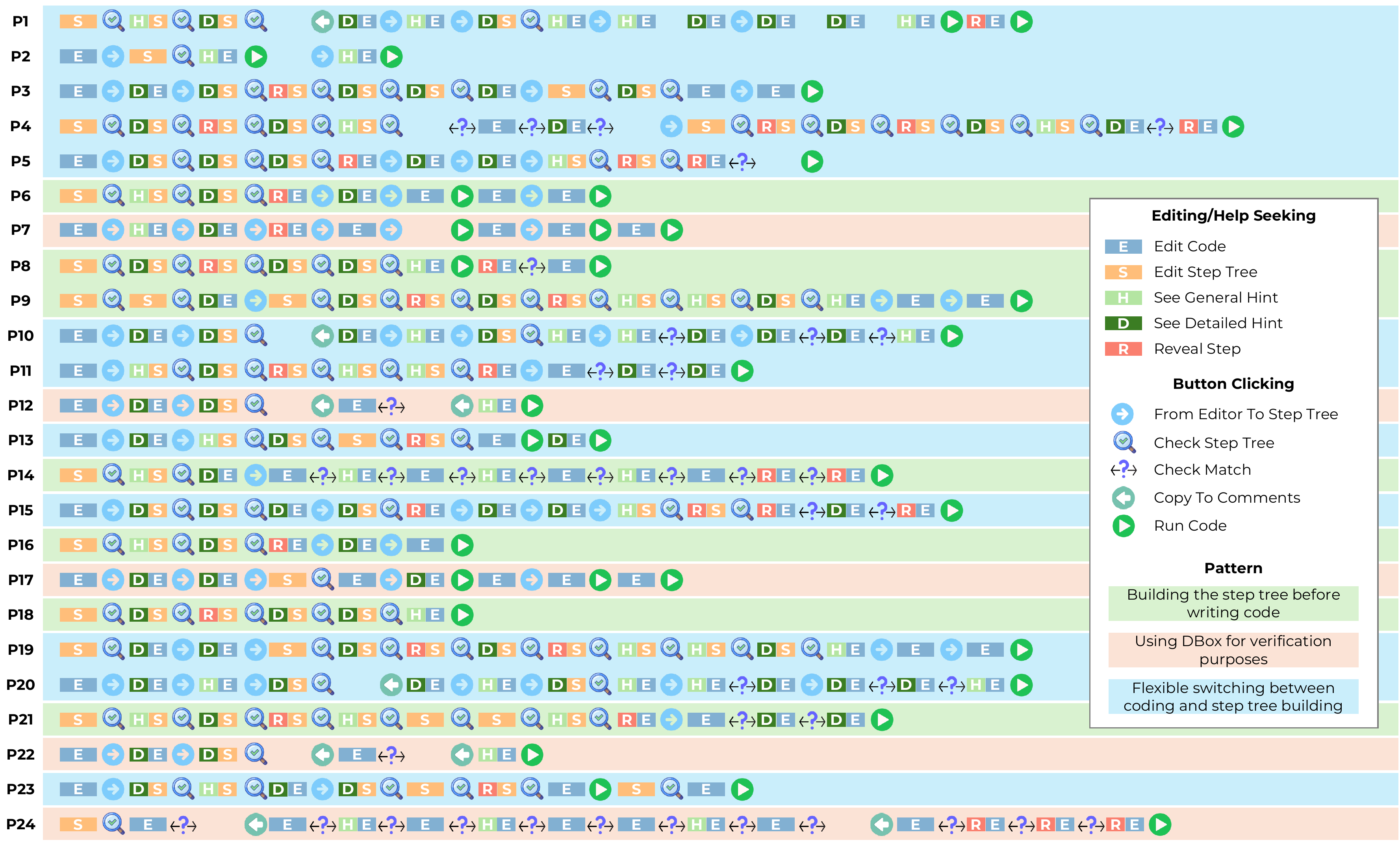}
	\caption{Participants' three distinct types of system usage, each represented by a different color. We analyzed participants' interactions, including code editing, step tree editing, help-seeking, and five types of button clicks.}
	\label{usage}
        \Description{}
\end{figure*}

\subsubsection{Effects on Learners' User Experience}
\label{Effects_on_UX}
As shown in Figure \ref{UX}, participants in the DBox condition found the learning process more mentally demanding ($Coef.$=1.125, $p$<0.01) and reported putting in more effort ($Coef.$=1.625, $p$<0.001) than those in the baseline condition. This aligns with our expectations, as DBox requires learners to independently construct a step tree rather than merely providing solutions. Interestingly, there were no significant differences in frustration levels between the two conditions ($Coef.$=-0.375, $p$=0.305), suggesting that while DBox encouraged independent thinking and led to some failed attempts, the process of building the step tree did not cause excessive frustration. 
\ms{However, we found a significant ordering effect on frustration ($Coef.$=1.917, $p$<0.01). 
Post-hoc analysis showed no significant difference between the two tools when DBox was used first ($t$=1.186, $p$=0.248), but participants reported significantly higher frustration with the baseline when it was used first ($t$=2.491, $p$<0.05). 
One possible explanation is the ``learning effect'' of soliciting assistance -- using DBox first better prepared participants to request targeted assistance from the baseline. 
This aligns how participants expressed frustration with the unsolicited assistance from the baseline. 
For example, P2 (using Copilot) commented, ``\emph{I typed a variable and hit [Tab], and Copilot suggested the entire code. However, the suggestion was different from what I had in mind, and I ended up spending time trying to understand it, which was frustrating.}''}

We expected participants to find DBox less easy to use due to its more complex operations, but participants' perceived ease of use ($Coef.$=0.750, $p$=0.063) did not differ significantly between conditions. Interviews revealed that in the baseline condition, participants often had to switch between the editor and solution pages or spend time crafting precise prompts for ChatGPT. As P6 noted, ``\emph{I had to explain the problem and my understanding to ChatGPT, which was quite complex. I didn't just want to copy its answer, so I constantly did line-by-line comparison between ChatGPT-provided code and my code.}'' Moreover, participants reported significantly higher satisfaction ($Coef.$=3.125, $p$<0.001) and a greater willingness to use DBox for future programming learning ($Coef.$=2.042, $p$<0.001).

\subsection{How do learners interact with Decomposition Box and perceive the usefulness of different features?}
\label{actual_use}

\subsubsection{Learners' Overall Usage Patterns}

During the user study, we tracked participants' interactions with DBox, focusing on key actions such as code editing, step tree editing, help-seeking, and five main button clicks (e.g., From Editor to Step Tree, Check Step Tree, Check Match, Copy to Comments, and Run Code). These interactions are visualized in Figure \ref{usage}, and we analyzed the patterns in combination with interview data.

Students adopted varying approaches when using DBox. Eleven began by constructing the step tree interactively, while thirteen started by writing code directly. We identified three distinct usage patterns:

\colorbox{color2}{\textbf{Type 1: Building the step tree before writing code}.} Some participants (P6, 8, 9, 14, 16, 18, and 21) focused on constructing the step tree first, iteratively checking and refining it before moving on to code implementation. This approach enabled them to write code efficiently once the structure was finalized. As P21 noted, ``\emph{I used natural language to express my thoughts and verify them, and after a few iterations, I recognized valid ideas and wrote the code myself.}'' P6 added, ``\emph{Writing code from scratch is more challenging for complex problems, so I prefer starting with the step tree.}''

\colorbox{color3}{\textbf{Type 2: Using DBox for verification}.} Some participants (P7, 12, 17, 22, and 24) used DBox primarily to verify their code. They wrote code first, then used the ``From Editor to Step Tree'' feature to check correctness and get hints. P17 explained, ``\emph{I usually solve problems by writing code first. Describing each step in natural language doesn’t feel natural for me.}'' Similarly, P22 stated, ``\emph{I know the general direction, so I write code first and use the tool to verify correctness or catch edge cases.}''

\colorbox{color1}{\textbf{Type 3: Flexible switching between the two modes.}} Students like P1-5, 10, 11, 13, 15, 19, 20, and 23 alternated between coding and step tree construction, adjusting their approach based on confidence, familiarity with specific steps, and real-time coding challenges. For example, P19 stated, ``\emph{If I'm confident in certain steps, I code first and then convert it to steps for verification. If unsure, I verify my thought process before coding.}'' P2 added, ``\emph{For familiar problems, I code first and refine it with the step tree. For new problems, I outline my thoughts and break down the steps to ensure accuracy before coding.}'' P10 shared, ``\emph{Initially, I felt confident, but when I got stuck, I refined my understanding of a step in the step tree before continuing with the code.}''

\begin{figure*}[htbp]
	\centering 
	\includegraphics[width=\linewidth]{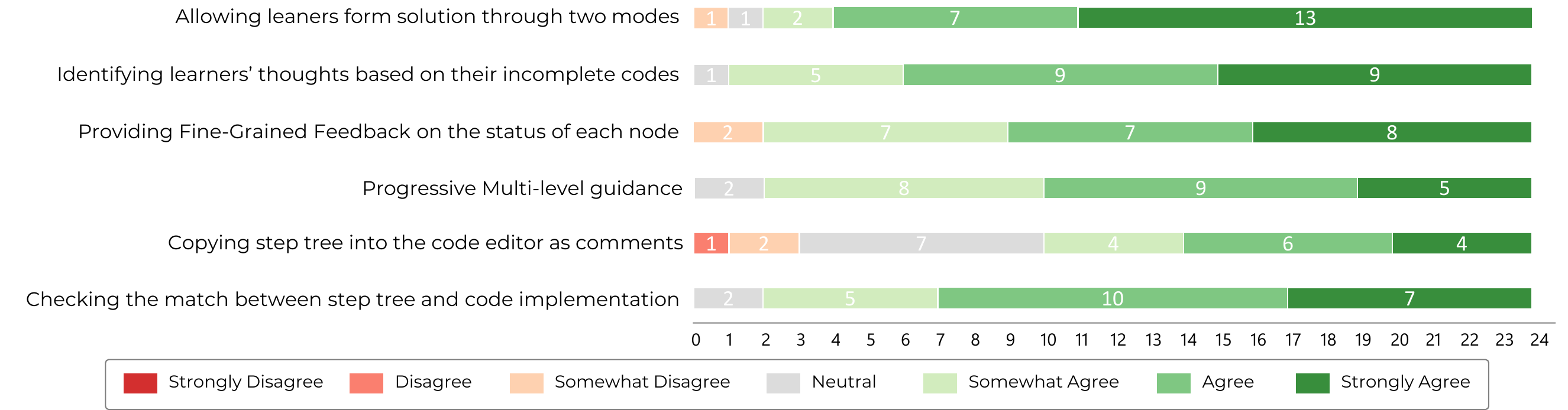}
	\caption{Participants rated the features of DBox based on their firsthand experience during the experiment. In the questionnaire, they provided feedback from a first-person perspective (e.g., ``I think [feature] is useful'', rated on a 7-point Likert scale).}
	\label{finalsurvey}
        \Description{}
\end{figure*}

\subsubsection{Hint Usage and Problem-Solving Approach}
\label{hintusage}
\ms{We tracked hint usage frequency among all 24 participants, recording a total of 164 hints triggered. Only 32 instances (19.5\%) involved the ``reveal sub-step'' feature, showing that students mainly relied on simpler hints and did not exploit the system by repeatedly making errors. This feature reveals only one sub-step from the user's incorrect or missing step—leaving the rest for them to solve independently—and can only be triggered after repeated struggle. This approach maximally preserves the students' independent problem-solving process. It strikes a balance between fostering independent thinking and preventing students from becoming permanently stuck, which could otherwise lead to frustration or a loss of motivation to learn.

We did a qualitative analysis which revealed varied problem-solving approaches adopted by participants in the problem-solving processes. For the ``Can Jump'' problem, tackled by 12 participants, we observed three approaches: Greedy (7 participants), Dynamic Programming (3), and Recursion (2). The other 12 participants addressed the ``Search in Rotated Sorted Array'' problem, using Binary Search (8), Two Pointers (2), Binary Tree (1), and Divide and Conquer (1). We observed that participants using the same approach still exhibited differences in reasoning and coding styles. Despite these variations, DBox effectively adapted its support to align with their individual styles and reasoning processes.
}

\subsubsection{Learners' Reactions to System Errors}
\ms{DBox, powered by the GPT-4o model, occasionally misjudged step statuses. During our user study, 24 participants triggered the ``check'' function (``Check Step Tree'' and ``From Editor to Step Tree'') 208 times, with 16 participants encountering 18 system errors (an 8.7\% error rate). Fourteen participants faced one error each, while two experienced two errors. These errors can be categorized into four types:

\begin{itemize}[leftmargin=0em] 

    \item \textbf{Type-1 (11 occurrences):} Misjudging correct steps as incorrect.  
    \begin{itemize}
        \item \textbf{Example:} A participant using a greedy approach stated, 
        \textit{``Use a greedy approach to minimize jumps.''}  
        GPT flagged this as incorrect due to insufficient detail on range expansion and jump counter updates. However, missing details does not necessarily mean the solution is incorrect.
    \end{itemize}

    \item \textbf{Type-2 (2 occurrences):} Flagging unnecessary steps as missing.  
    \begin{itemize}
        \item \textbf{Example:} GPT incorrectly required a check for single-element arrays, though the solution worked without it.  
    \end{itemize}

    \item \textbf{Type-3 (3 occurrences):} Overlooking subtle mistakes in seemingly correct steps.  
    \begin{itemize}
        \item \textbf{Example:} A participant's wrote a step ``Iterate through the array. If at any index \texttt{i}, \texttt{maxReach < i}, return false'', leading to errors with unreachable indices.  
        GPT failed to detect this error.  
    \end{itemize}

    \item \textbf{Type-4 (2 occurrences):} Missing crucial steps while DBox marking solutions as complete.  
    \begin{itemize}
        \item \textbf{Example:} A participant omitted a final return statement (\texttt{return true}), GPT still judged the solution as correct.  
    \end{itemize}

\end{itemize}
}


We then analyzed whether participants successfully identified the system errors and how they reacted to the errors. We found that all the 16 students who encountered system errors recognized these errors after one or more attempts. Since DBox evaluates rather than generates content, students remained confident in verifying their own work, which minimized over-reliance on the system. However, incorrect judgments, particularly when correct steps were flagged as wrong, could disrupt their thought process. The impact of these errors largely depended on the student's confidence. As P15 noted, “\emph{If I know it's wrong, I ignore the message. If I'm unsure, I might follow the hint.}''

Students adopted various strategies to deal with system errors. (1) \textbf{Ignoring the error} (8 participants): Some simply moved on after recognizing a misjudgment. (2) \textbf{Focusing on the hint, not the status} (6 participants): Even if the status evaluation was incorrect, participants often found the hints useful. \ms{For example, participants wrote very brief steps, which GPT flagged as incorrect. The hints prompted them to consider important details, and despite the incorrect judgment, participants found the hints very useful.} (3) \textbf{Running the code to verify} (12 participants): Many used the ``Run'' button to test their hypotheses. As P2 explained, ``\emph{If my code works, I stick with my approach. If it fails, I reconsider the system's judgment.}'' (4) \textbf{Rechecking the step} (10 participants): Some participants revisited a step to refresh its status. As P3 mentioned, ``\emph{Even if the system is wrong, rechecking helps me identify potential issues.}''

Though system errors are unavoidable at the current stage, DBox’s design allows learners to quickly recognize and manage them. Future improvements should aim to minimize disruptions caused by system inaccuracies.

\subsubsection{Learners' Ratings of Different Features of DBox}
DBox offers a range of features designed to enhance algorithmic programming learning, most of which participants found helpful, as illustrated in Figure \ref{finalsurvey}. Students particularly appreciated the flexibility to either write code directly or build a step tree, and they valued DBox's ability to infer their thought process from the code. The interactive step tree and the fine-grained correctness assessment were also well-received, although two participants found it somewhat cumbersome. The progressive, multi-level hint system was praised by 22 participants. Additionally, the feature that checks the alignment between the step tree and the code implementation was seen as highly beneficial. However, some participants felt that the ability to paste the step tree as comments into the editor was unnecessary, since the step tree and code editor could already be viewed side by side. This feature could be even more useful if DBox were developed as a plugin in the future.

\section{Discussion}

In this paper, we adopted a learner-centered design approach, beginning with a formative study to identify students' challenges with existing tools. Based on these insights, we developed DBox, a tool that scaffolds students in breaking problems into smaller parts and provides personalized, adaptive support. Our user study demonstrated that DBox improved learners' performance on similar algorithmic problems, increased perceived learning gains, and fostered greater cognitive engagement, achievement, and satisfaction. In this section, we discuss design implications and generalizability based on our key findings.

\ms{
\subsection{Chaining Learners' Thoughts with Visualized Structured UI Components}

Decomposition requires students to effectively organize their thoughts. While visual elements are known to promote structured thinking and support mental model construction \cite{mcdougall2001effects, liu2010mental}, our formative and user studies revealed shortcomings in existing tools like LeetCode and ChatGPT, which rely on textual representations without adequately supporting structured mental models. In contrast, DBox uses an interactive step tree to visually organize learners' thoughts. This feature was praised by 22 of 24 participants for enhancing algorithmic thinking, serving as a progress tracker, and providing value even without AI assistance.

DBox's interactive step tree and tree-based scaffolding demonstrate the broader potential of intelligent tutoring systems (ITS) to promote active learning and self-regulated problem-solving in fields requiring problem decomposition. Similar principles could benefit STEM education, such as physics or engineering, by externalizing abstract concepts and facilitating multi-step problem-solving. Additionally, progress-tracking visual components may inspire designs for professional training tools in areas like medical diagnostics or software engineering.

\subsection{Promoting Independent Thinking and Active Decomposition Learning}

\subsubsection{\textbf{Transforming Learners from Passive Readers to Active Thinkers}}

Many coding tools provide direct answers or solutions \cite{kazemitabaar2023novices, phung2023generating}, which, while efficient, often bypass opportunities to develop critical problem-solving skills. In contrast, DBox cultivates students' decomposition abilities through structured scaffolding, fostering critical thinking and self-regulated learning in line with learning by doing \cite{anzai1979theory} and constructivist principles \cite{tobias2009constructivist}.

To strengthen decomposition skills, DBox first encourages students to develop their own decomposition strategies by coding or building a step tree from scratch. While DBox can generate parts of a step tree from a student's existing code, these steps are derived from the learner's own reasoning, with DBox acting solely as a modality converter. Besides, DBox provides feedback on tree node statuses, identifying potential errors or missing steps without directly showing the correct answer, challenging students to critically evaluate and refine their decomposition plans.

DBox's scaffolded hint system further supports decomposition skill development by providing adaptive guidance tailored to the student’s progress without overwhelming them. All hints are based on the learner's current decomposition skeleton, with the most detailed hint—``reveal substep''—triggered only after repeated attempts and struggles. Notably, even the most detailed hints prompt only one substep, requiring students to complete the rest independently. As shown in Sec \ref{hintusage}, only 19\% of hints are this detailed, with students primarily relying on simpler, thought-provoking question hints. This scaffolded support system balances guidance and independent thinking, keeping students engaged during challenges without compromising their ability to independently decompose problems \cite{kinnunen2006students}.

Based on these findings, we recommend fostering active problem-solving by shifting students from passive content consumption to active solution creation. Designers could adopt layered scaffolding, starting with minimal guidance and increasing support as needed, to help students progressively master decomposition skills while maintaining confidence and avoiding frustration. Additionally, adaptive learning techniques, such as real-time feedback and progress tracking, can further tailor the support to individual decomposition barriers, encouraging deeper engagement with decomposition tasks. Moreover, designers could integrate metacognitive strategies, such as encouraging students to articulate or reflect on their decomposition approaches, to further enhance critical thinking and foster habits of independent thinking.

\subsubsection{\textbf{Choice of Scaffolding: Balancing Independent Problem-Solving and Efforts}}

Scaffolding involves providing tailored support to help learners accomplish tasks they cannot yet complete independently \cite{kim2011scaffolding, tobias2009constructivist}. Broadly, scaffolding strategies fall into two categories \cite{van2010scaffolding}: (1) gradually reducing assistance as learners gain proficiency, and (2) encouraging independent problem-solving while offering incremental support to address challenges. DBox adopts the second approach, emphasizing independent thinking and encouraging learners to actively decompose problems \cite{zimmerman2013theories}. While our scaffolding strategies successfully enhanced critical thinking, satisfaction, and perceived usefulness, they also led to increased cognitive effort (Sec. \ref{Effects_on_UX}). This tradeoff underscores the importance of carefully balancing cognitive effort with the promotion of independent thinking.

Future designs could incorporate adaptive scaffolding that adjusts support dynamically based on learner proficiency, reducing unnecessary effort in areas where students have demonstrated competence. Additionally, while incremental scaffolding was effective for algorithmic problem-solving, tailoring strategies to different educational contexts could enhance their applicability in diverse domains. Such adaptive, context-specific approaches could further optimize the balance between support and independence in learning environments.

\subsection{Supporting Personalized Algorithmic Programming Learning}

\subsubsection{\textbf{Prioritizing Learners' Own Solutions Over Optimality}}

Algorithmic problems often have multiple solutions with varying time and space complexities. DBox prioritizes independent exploration by supporting learners' strategies rather than steering them toward a single ``optimal'' solution. Using LLM-driven prompts, it evaluates and guides each step based on the learner's reasoning, preserving their step decomposition and respecting their input—even when errors occur. While some solutions may not be the most efficient, this approach fosters autonomy by aligning feedback with learners’ thought processes instead of enforcing rigid standards.

Our user study showed that this approach improves learning outcomes and is well-received by students. We recommend designing systems that respect personalized problem-solving strategies by aligning feedback with learners' reasoning while allowing for diverse approaches. Designers should balance flexibility and rigor, using prompts and interfaces that support varied strategies while gently guiding learners toward effective solutions.

\subsubsection{\textbf{Catering to Individual Learning Styles and Contextual Needs}}

DBox accommodates diverse problem-solving approaches with two input modes: coding and natural language descriptions. Each mode offers distinct advantages tailored to different learners, stages, and situations. Learners can switch seamlessly between modes, with progress automatically synced across the interface. Features such as verifying code-step alignment ensure strong integration between modes.

Our findings reveal that this flexibility enhances user experience. Participant interaction logs and interviews revealed three usage patterns, highlighting that each mode fits different needs: code mode works well for students with a clear and detailed problem-solving plan already, while the step tree with natural language descriptions helps less experienced students with only a basic idea who are not ready to write code directly, boosting their confidence.

We argue there is no universal “best” mode for programming education—each has unique benefits depending on the learner habits, expertise, and context. Future tools should provide flexibility, like DBox, or use adaptive algorithms to recommend modes based on user needs and context. This flexibility highlights the importance of designing educational tools that accommodate varying levels of expertise and problem-solving styles, which can be generalized to other domains requiring personalized learning \cite{bernacki2021systematic}.

\subsection{Appropriate Usage of LLMs for Supporting Algorithmic Programming Learning}

\subsubsection{\textbf{Caution About LLM Errors}}

Although LLMs have shown strong performance in coding tasks \cite{finnie2023my, leinonen2023using}, they remain prone to errors. Our technical evaluation and user study revealed that even with comprehensive context—such as problem statements, user code, and natural language steps—LLM sometimes misinterprets user descriptions. These errors likely arise from discrepancies between the natural language used by students and the formal, precise language the LLM was trained on, which is primarily sourced from web-based code and comments \cite{liu2023wants}.

Such misinterpretations can hinder learning by causing confusion or frustration. While future improvements to training data and GPT versions may mitigate these issues, design strategies can help address them. \textbf{First}, LLMs should avoid giving direct solutions and instead focus on fostering active problem-solving through explanations and hints. \textbf{Second}, feedback could be paired with interactive features, like a ``Run Code'' option, allowing students to validate their reasoning. \textbf{Third}, simple tutorials could teach users how to phrase their descriptions more clearly, improving LLM's understanding. Additionally, future tools could integrate a ``Language Enhancement'' feature to suggest improvements or assess the clarity of descriptions, aiding LLM in accurately capturing user intent. Most importantly, we recommend designers prioritize technical feasibility, such as conducting rigorous evaluations like ours, before fully integrating LLMs into programming learning tools.
}

\subsubsection{\textbf{Learner-LLM Co-Decomposition of Solutions: Learner as Leader, LLM as Aid}}

A central feature of DBox is the construction of a step tree, where students break solutions into steps and sub-steps. The LLM supports this by mapping code to step descriptions, evaluating them, and offering hints. However, students maintain full control, deciding how to decompose problems and define each step, fostering independent thinking. The LLM acts solely as an aid, using a scaffolding approach to support the development of learners' Zone of Proximal Development (ZPD) \cite{chaiklin2003zone}. Unlike tools like ChatGPT or Copilot that dominate problem-solving, DBox fosters deeper cognitive engagement. Students reported greater accomplishment and found this approach more effective for learning.

This contrasts with existing human-AI collaboration paradigms in non-educational scenarios where AI usually suggest options, leaving final decisions to users \cite{dang2023choice, gao2024collabcoder, gebreegziabher2023patat, ma2019smarteye, ma2022glancee}, such as in human-AI decision-making \cite{ma2023should, ma2024towards, ma2024you}. Some educational tools, like Jin et al. \cite{jin2024teach}, use LLMs to generate solutions for students to evaluate, which aids in syntax learning but such ``LLM-generate then learner-evaluate'' approach is less effective for algorithmic problem-solving, where constructing solutions is key. Just evaluating LLM-generated contents can place a cognitive anchor on learners \cite{furnham2011literature}, limiting independent thinking and creativity. Thus, task allocation between humans and AI should align with the educational context (e.g., whether it is basic knowledge/concept learning or higher-level creative thinking). Future LLM-based educational tools should carefully define the division of roles between LLMs and learners, tailoring it to specific learning contexts and goals.

\subsection{Limitations and Future Work}

This study has several limitations. \emph{First}, we tested DBox's effectiveness on only two problem types; future work should examine a broader range of algorithms. \emph{Second}, participants engaged in just one learning session per condition due to time constraints, whereas mastering algorithmic problems typically requires extended practice. Longitudinal studies should explore how DBox supports skill development over time, including changes in mental models and skill retention. \emph{Third}, we assessed learning gains based on correctness in a test session using similar learning and test problems. Future research should evaluate knowledge transfer to less similar problems. Due to time constraints, we conducted a single post-test rather than a pre-post comparison. While pre-test expertise filtering and randomization minimized prior familiarity effects, a more rigorous pre-post design would yield more accurate learning gain measurements. Looking ahead, we plan to release DBox as a Chrome plugin for integration with existing coding platforms, enabling large-scale field studies. This will allow for the collection of long-term usage data and periodic surveys to identify usage patterns and learning experiences over time.

\section{Conclusion}
In this paper, we introduced DBox, an interactive tool designed to help learners decompose algorithmic programming problems by supporting both solution formation and implementation. Featuring an intuitive tree-like box widget, DBox accepts input in both code and natural language, fostering independent problem-solving while its step tree structure helps learners develop structured mental models. It provides step-level feedback and layered guidance without compromising learner autonomy.
Our user study showed that DBox significantly improved learning outcomes, cognitive engagement, and critical thinking, with students reporting a greater sense of achievement and finding the support highly effective. Additionally, we identified three key usage patterns, highlighting the importance of accommodating individual problem-solving styles. Moreover, our findings suggest that the learner-LLM co-decomposition approach fosters independent thinking while providing meaningful guidance, even with occasional imperfections.
We hope the insights from our system design will inspire future research on integrating LLMs into educational tools for programming learning.

\begin{acks}
This research was supported by the Dieter Schwarz Stiftung Foundation, ETH Foundation, and in part by the EdUHK-HKUST Joint Centre for Artificial Intelligence (JC\_AI) research scheme: Grant No. FB454.
\end{acks}

\bibliographystyle{ACM-Reference-Format}
\bibliography{sample-base}


\begin{thebibliography}{85}


\ifx \showCODEN    \undefined \def \showCODEN     #1{\unskip}     \fi
\ifx \showDOI      \undefined \def \showDOI       #1{#1}\fi
\ifx \showISBNx    \undefined \def \showISBNx     #1{\unskip}     \fi
\ifx \showISBNxiii \undefined \def \showISBNxiii  #1{\unskip}     \fi
\ifx \showISSN     \undefined \def \showISSN      #1{\unskip}     \fi
\ifx \showLCCN     \undefined \def \showLCCN      #1{\unskip}     \fi
\ifx \shownote     \undefined \def \shownote      #1{#1}          \fi
\ifx \showarticletitle \undefined \def \showarticletitle #1{#1}   \fi
\ifx \showURL      \undefined \def \showURL       {\relax}        \fi
\providecommand\bibfield[2]{#2}
\providecommand\bibinfo[2]{#2}
\providecommand\natexlab[1]{#1}
\providecommand\showeprint[2][]{arXiv:#2}

\bibitem[\protect\citeauthoryear{Academy}{Academy}{2024}]%
        {codetutor}
\bibfield{author}{\bibinfo{person}{Khan Academy}.} \bibinfo{year}{2024}\natexlab{}.
\newblock \bibinfo{title}{Code Tutor}.
\newblock
\newblock
\urldef\tempurl%
\url{https://chatgpt.com/g/g-HxPrv1p8v-code-tutor}
\showURL{%
\tempurl}
\newblock
\shownote{Accessed: September 10, 2024.}


\bibitem[\protect\citeauthoryear{Achiam, Adler, Agarwal, Ahmad, Akkaya, Aleman, Almeida, Altenschmidt, Altman, Anadkat, et~al\mbox{.}}{Achiam et~al\mbox{.}}{2023}]%
        {achiam2023gpt}
\bibfield{author}{\bibinfo{person}{Josh Achiam}, \bibinfo{person}{Steven Adler}, \bibinfo{person}{Sandhini Agarwal}, \bibinfo{person}{Lama Ahmad}, \bibinfo{person}{Ilge Akkaya}, \bibinfo{person}{Florencia~Leoni Aleman}, \bibinfo{person}{Diogo Almeida}, \bibinfo{person}{Janko Altenschmidt}, \bibinfo{person}{Sam Altman}, \bibinfo{person}{Shyamal Anadkat}, {et~al\mbox{.}}} \bibinfo{year}{2023}\natexlab{}.
\newblock \showarticletitle{Gpt-4 technical report}.
\newblock \bibinfo{journal}{\emph{arXiv preprint arXiv:2303.08774}} (\bibinfo{year}{2023}).
\newblock


\bibitem[\protect\citeauthoryear{Aleven and Koedinger}{Aleven and Koedinger}{2002}]%
        {aleven2002effective}
\bibfield{author}{\bibinfo{person}{Vincent~AWMM Aleven} {and} \bibinfo{person}{Kenneth~R Koedinger}.} \bibinfo{year}{2002}\natexlab{}.
\newblock \showarticletitle{An effective metacognitive strategy: Learning by doing and explaining with a computer-based cognitive tutor}.
\newblock \bibinfo{journal}{\emph{Cognitive science}} \bibinfo{volume}{26}, \bibinfo{number}{2} (\bibinfo{year}{2002}), \bibinfo{pages}{147--179}.
\newblock


\bibitem[\protect\citeauthoryear{Angeli, Voogt, Fluck, Webb, Cox, Malyn-Smith, and Zagami}{Angeli et~al\mbox{.}}{2016}]%
        {angeli2016k}
\bibfield{author}{\bibinfo{person}{Charoula Angeli}, \bibinfo{person}{Joke Voogt}, \bibinfo{person}{Andrew Fluck}, \bibinfo{person}{Mary Webb}, \bibinfo{person}{Margaret Cox}, \bibinfo{person}{Joyce Malyn-Smith}, {and} \bibinfo{person}{Jason Zagami}.} \bibinfo{year}{2016}\natexlab{}.
\newblock \showarticletitle{A K-6 computational thinking curriculum framework: Implications for teacher knowledge}.
\newblock \bibinfo{journal}{\emph{Journal of Educational Technology \& Society}} \bibinfo{volume}{19}, \bibinfo{number}{3} (\bibinfo{year}{2016}), \bibinfo{pages}{47--57}.
\newblock


\bibitem[\protect\citeauthoryear{Anzai and Simon}{Anzai and Simon}{1979}]%
        {anzai1979theory}
\bibfield{author}{\bibinfo{person}{Yuichiro Anzai} {and} \bibinfo{person}{Herbert~A Simon}.} \bibinfo{year}{1979}\natexlab{}.
\newblock \showarticletitle{The theory of learning by doing.}
\newblock \bibinfo{journal}{\emph{Psychological review}} \bibinfo{volume}{86}, \bibinfo{number}{2} (\bibinfo{year}{1979}), \bibinfo{pages}{124}.
\newblock


\bibitem[\protect\citeauthoryear{Backhouse}{Backhouse}{2011}]%
        {backhouse2011algorithmic}
\bibfield{author}{\bibinfo{person}{Roland Backhouse}.} \bibinfo{year}{2011}\natexlab{}.
\newblock \bibinfo{booktitle}{\emph{Algorithmic problem solving}}.
\newblock \bibinfo{publisher}{John Wiley \& Sons}.
\newblock


\bibitem[\protect\citeauthoryear{Bangor, Kortum, and Miller}{Bangor et~al\mbox{.}}{2008}]%
        {bangor2008empirical}
\bibfield{author}{\bibinfo{person}{Aaron Bangor}, \bibinfo{person}{Philip~T Kortum}, {and} \bibinfo{person}{James~T Miller}.} \bibinfo{year}{2008}\natexlab{}.
\newblock \showarticletitle{An empirical evaluation of the system usability scale}.
\newblock \bibinfo{journal}{\emph{Intl. Journal of Human--Computer Interaction}} \bibinfo{volume}{24}, \bibinfo{number}{6} (\bibinfo{year}{2008}), \bibinfo{pages}{574--594}.
\newblock


\bibitem[\protect\citeauthoryear{Bernacki, Greene, and Lobczowski}{Bernacki et~al\mbox{.}}{2021}]%
        {bernacki2021systematic}
\bibfield{author}{\bibinfo{person}{Matthew~L Bernacki}, \bibinfo{person}{Meghan~J Greene}, {and} \bibinfo{person}{Nikki~G Lobczowski}.} \bibinfo{year}{2021}\natexlab{}.
\newblock \showarticletitle{A systematic review of research on personalized learning: Personalized by whom, to what, how, and for what purpose (s)?}
\newblock \bibinfo{journal}{\emph{Educational Psychology Review}} \bibinfo{volume}{33}, \bibinfo{number}{4} (\bibinfo{year}{2021}), \bibinfo{pages}{1675--1715}.
\newblock


\bibitem[\protect\citeauthoryear{Chaiklin et~al\mbox{.}}{Chaiklin et~al\mbox{.}}{2003}]%
        {chaiklin2003zone}
\bibfield{author}{\bibinfo{person}{Seth Chaiklin} {et~al\mbox{.}}} \bibinfo{year}{2003}\natexlab{}.
\newblock \showarticletitle{The zone of proximal development in Vygotsky’s analysis of learning and instruction}.
\newblock \bibinfo{journal}{\emph{Vygotsky’s educational theory in cultural context}} \bibinfo{volume}{1}, \bibinfo{number}{2} (\bibinfo{year}{2003}), \bibinfo{pages}{39--64}.
\newblock


\bibitem[\protect\citeauthoryear{Cole, John-Steiner, Scribner, and Souberman}{Cole et~al\mbox{.}}{1978}]%
        {cole1978mind}
\bibfield{author}{\bibinfo{person}{Michael Cole}, \bibinfo{person}{Vera John-Steiner}, \bibinfo{person}{Sylvia Scribner}, {and} \bibinfo{person}{Ellen Souberman}.} \bibinfo{year}{1978}\natexlab{}.
\newblock \showarticletitle{Mind in society}.
\newblock \bibinfo{journal}{\emph{Mind in society the development of higher psychological processes. Cambridge, MA: Harvard University Press}} (\bibinfo{year}{1978}).
\newblock


\bibitem[\protect\citeauthoryear{Conati and Vanlehn}{Conati and Vanlehn}{2000}]%
        {conati2000toward}
\bibfield{author}{\bibinfo{person}{Cristina Conati} {and} \bibinfo{person}{Kurt Vanlehn}.} \bibinfo{year}{2000}\natexlab{}.
\newblock \showarticletitle{Toward computer-based support of meta-cognitive skills: A computational framework to coach self-explanation}.
\newblock \bibinfo{journal}{\emph{International Journal of Artificial Intelligence in Education}}  \bibinfo{volume}{11} (\bibinfo{year}{2000}), \bibinfo{pages}{389--415}.
\newblock


\bibitem[\protect\citeauthoryear{Cunningham, Ericson, Agrawal~Bejarano, and Guzdial}{Cunningham et~al\mbox{.}}{2021}]%
        {cunningham2021avoiding}
\bibfield{author}{\bibinfo{person}{Kathryn Cunningham}, \bibinfo{person}{Barbara~J Ericson}, \bibinfo{person}{Rahul Agrawal~Bejarano}, {and} \bibinfo{person}{Mark Guzdial}.} \bibinfo{year}{2021}\natexlab{}.
\newblock \showarticletitle{Avoiding the Turing tarpit: Learning conversational programming by starting from code’s purpose}. In \bibinfo{booktitle}{\emph{Proceedings of the 2021 CHI Conference on Human Factors in Computing Systems}}. \bibinfo{pages}{1--15}.
\newblock


\bibitem[\protect\citeauthoryear{Dang, Goller, Lehmann, and Buschek}{Dang et~al\mbox{.}}{2023}]%
        {dang2023choice}
\bibfield{author}{\bibinfo{person}{Hai Dang}, \bibinfo{person}{Sven Goller}, \bibinfo{person}{Florian Lehmann}, {and} \bibinfo{person}{Daniel Buschek}.} \bibinfo{year}{2023}\natexlab{}.
\newblock \showarticletitle{Choice over control: How users write with large language models using diegetic and non-diegetic prompting}. In \bibinfo{booktitle}{\emph{Proceedings of the 2023 CHI Conference on Human Factors in Computing Systems}}. \bibinfo{pages}{1--17}.
\newblock


\bibitem[\protect\citeauthoryear{Davis}{Davis}{1989}]%
        {davis1989perceived}
\bibfield{author}{\bibinfo{person}{Fred~D Davis}.} \bibinfo{year}{1989}\natexlab{}.
\newblock \showarticletitle{Perceived usefulness, perceived ease of use, and user acceptance of information technology}.
\newblock \bibinfo{journal}{\emph{MIS quarterly}} (\bibinfo{year}{1989}), \bibinfo{pages}{319--340}.
\newblock


\bibitem[\protect\citeauthoryear{Delen, Liew, and Willson}{Delen et~al\mbox{.}}{2014}]%
        {delen2014effects}
\bibfield{author}{\bibinfo{person}{Erhan Delen}, \bibinfo{person}{Jeffrey Liew}, {and} \bibinfo{person}{Victor Willson}.} \bibinfo{year}{2014}\natexlab{}.
\newblock \showarticletitle{Effects of interactivity and instructional scaffolding on learning: Self-regulation in online video-based environments}.
\newblock \bibinfo{journal}{\emph{Computers \& Education}}  \bibinfo{volume}{78} (\bibinfo{year}{2014}), \bibinfo{pages}{312--320}.
\newblock


\bibitem[\protect\citeauthoryear{Denny, Kumar, and Giacaman}{Denny et~al\mbox{.}}{2023}]%
        {denny2023conversing}
\bibfield{author}{\bibinfo{person}{Paul Denny}, \bibinfo{person}{Viraj Kumar}, {and} \bibinfo{person}{Nasser Giacaman}.} \bibinfo{year}{2023}\natexlab{}.
\newblock \showarticletitle{Conversing with copilot: Exploring prompt engineering for solving cs1 problems using natural language}. In \bibinfo{booktitle}{\emph{Proceedings of the 54th ACM Technical Symposium on Computer Science Education V. 1}}. \bibinfo{pages}{1136--1142}.
\newblock


\bibitem[\protect\citeauthoryear{Denny, Sarsa, Hellas, and Leinonen}{Denny et~al\mbox{.}}{2022}]%
        {denny2022robosourcing}
\bibfield{author}{\bibinfo{person}{Paul Denny}, \bibinfo{person}{Sami Sarsa}, \bibinfo{person}{Arto Hellas}, {and} \bibinfo{person}{Juho Leinonen}.} \bibinfo{year}{2022}\natexlab{}.
\newblock \showarticletitle{Robosourcing Educational Resources--Leveraging Large Language Models for Learnersourcing}.
\newblock \bibinfo{journal}{\emph{arXiv preprint arXiv:2211.04715}} (\bibinfo{year}{2022}).
\newblock


\bibitem[\protect\citeauthoryear{Faul, Erdfelder, Buchner, and Lang}{Faul et~al\mbox{.}}{2009}]%
        {faul2009statistical}
\bibfield{author}{\bibinfo{person}{Franz Faul}, \bibinfo{person}{Edgar Erdfelder}, \bibinfo{person}{Axel Buchner}, {and} \bibinfo{person}{Albert-Georg Lang}.} \bibinfo{year}{2009}\natexlab{}.
\newblock \showarticletitle{Statistical power analyses using G* Power 3.1: Tests for correlation and regression analyses}.
\newblock \bibinfo{journal}{\emph{Behavior research methods}} \bibinfo{volume}{41}, \bibinfo{number}{4} (\bibinfo{year}{2009}), \bibinfo{pages}{1149--1160}.
\newblock


\bibitem[\protect\citeauthoryear{Finnie-Ansley, Denny, Becker, Luxton-Reilly, and Prather}{Finnie-Ansley et~al\mbox{.}}{2022}]%
        {finnie2022robots}
\bibfield{author}{\bibinfo{person}{James Finnie-Ansley}, \bibinfo{person}{Paul Denny}, \bibinfo{person}{Brett~A Becker}, \bibinfo{person}{Andrew Luxton-Reilly}, {and} \bibinfo{person}{James Prather}.} \bibinfo{year}{2022}\natexlab{}.
\newblock \showarticletitle{The robots are coming: Exploring the implications of openai codex on introductory programming}. In \bibinfo{booktitle}{\emph{Proceedings of the 24th Australasian Computing Education Conference}}. \bibinfo{pages}{10--19}.
\newblock


\bibitem[\protect\citeauthoryear{Finnie-Ansley, Denny, Luxton-Reilly, Santos, Prather, and Becker}{Finnie-Ansley et~al\mbox{.}}{2023}]%
        {finnie2023my}
\bibfield{author}{\bibinfo{person}{James Finnie-Ansley}, \bibinfo{person}{Paul Denny}, \bibinfo{person}{Andrew Luxton-Reilly}, \bibinfo{person}{Eddie~Antonio Santos}, \bibinfo{person}{James Prather}, {and} \bibinfo{person}{Brett~A Becker}.} \bibinfo{year}{2023}\natexlab{}.
\newblock \showarticletitle{My ai wants to know if this will be on the exam: Testing openai’s codex on cs2 programming exercises}. In \bibinfo{booktitle}{\emph{Proceedings of the 25th Australasian Computing Education Conference}}. \bibinfo{pages}{97--104}.
\newblock


\bibitem[\protect\citeauthoryear{Furnham and Boo}{Furnham and Boo}{2011}]%
        {furnham2011literature}
\bibfield{author}{\bibinfo{person}{Adrian Furnham} {and} \bibinfo{person}{Hua~Chu Boo}.} \bibinfo{year}{2011}\natexlab{}.
\newblock \showarticletitle{A literature review of the anchoring effect}.
\newblock \bibinfo{journal}{\emph{The journal of socio-economics}} \bibinfo{volume}{40}, \bibinfo{number}{1} (\bibinfo{year}{2011}), \bibinfo{pages}{35--42}.
\newblock


\bibitem[\protect\citeauthoryear{Gao, Guo, Lim, Zhang, Zhang, Li, and Perrault}{Gao et~al\mbox{.}}{2024}]%
        {gao2024collabcoder}
\bibfield{author}{\bibinfo{person}{Jie Gao}, \bibinfo{person}{Yuchen Guo}, \bibinfo{person}{Gionnieve Lim}, \bibinfo{person}{Tianqin Zhang}, \bibinfo{person}{Zheng Zhang}, \bibinfo{person}{Toby Jia-Jun Li}, {and} \bibinfo{person}{Simon~Tangi Perrault}.} \bibinfo{year}{2024}\natexlab{}.
\newblock \showarticletitle{CollabCoder: a lower-barrier, rigorous workflow for inductive collaborative qualitative analysis with large language models}. In \bibinfo{booktitle}{\emph{Proceedings of the CHI Conference on Human Factors in Computing Systems}}. \bibinfo{pages}{1--29}.
\newblock


\bibitem[\protect\citeauthoryear{Gebreegziabher, Zhang, Tang, Meng, Glassman, and Li}{Gebreegziabher et~al\mbox{.}}{2023}]%
        {gebreegziabher2023patat}
\bibfield{author}{\bibinfo{person}{Simret~Araya Gebreegziabher}, \bibinfo{person}{Zheng Zhang}, \bibinfo{person}{Xiaohang Tang}, \bibinfo{person}{Yihao Meng}, \bibinfo{person}{Elena~L Glassman}, {and} \bibinfo{person}{Toby Jia-Jun Li}.} \bibinfo{year}{2023}\natexlab{}.
\newblock \showarticletitle{Patat: Human-ai collaborative qualitative coding with explainable interactive rule synthesis}. In \bibinfo{booktitle}{\emph{Proceedings of the 2023 CHI Conference on Human Factors in Computing Systems}}. \bibinfo{pages}{1--19}.
\newblock


\bibitem[\protect\citeauthoryear{Hart}{Hart}{2006}]%
        {hart2006nasa}
\bibfield{author}{\bibinfo{person}{Sandra~G Hart}.} \bibinfo{year}{2006}\natexlab{}.
\newblock \showarticletitle{NASA-task load index (NASA-TLX); 20 years later}. In \bibinfo{booktitle}{\emph{Proceedings of the human factors and ergonomics society annual meeting}}, Vol.~\bibinfo{volume}{50}. Sage publications Sage CA: Los Angeles, CA, \bibinfo{pages}{904--908}.
\newblock


\bibitem[\protect\citeauthoryear{Hendriana, Johanto, and Sumarmo}{Hendriana et~al\mbox{.}}{2018}]%
        {hendriana2018role}
\bibfield{author}{\bibinfo{person}{Heris Hendriana}, \bibinfo{person}{Tri Johanto}, {and} \bibinfo{person}{Utari Sumarmo}.} \bibinfo{year}{2018}\natexlab{}.
\newblock \showarticletitle{The Role of Problem-Based Learning to Improve Students' Mathematical Problem-Solving Ability and Self Confidence.}
\newblock \bibinfo{journal}{\emph{Journal on Mathematics Education}} \bibinfo{volume}{9}, \bibinfo{number}{2} (\bibinfo{year}{2018}), \bibinfo{pages}{291--300}.
\newblock


\bibitem[\protect\citeauthoryear{Hmelo-Silver, Duncan, and Chinn}{Hmelo-Silver et~al\mbox{.}}{2007}]%
        {hmelo2007scaffolding}
\bibfield{author}{\bibinfo{person}{Cindy~E Hmelo-Silver}, \bibinfo{person}{Ravit~Golan Duncan}, {and} \bibinfo{person}{Clark~A Chinn}.} \bibinfo{year}{2007}\natexlab{}.
\newblock \showarticletitle{Scaffolding and achievement in problem-based and inquiry learning: a response to Kirschner, Sweller, and}.
\newblock \bibinfo{journal}{\emph{Educational psychologist}} \bibinfo{volume}{42}, \bibinfo{number}{2} (\bibinfo{year}{2007}), \bibinfo{pages}{99--107}.
\newblock


\bibitem[\protect\citeauthoryear{Holden and Karsh}{Holden and Karsh}{2010}]%
        {holden2010technology}
\bibfield{author}{\bibinfo{person}{Richard~J Holden} {and} \bibinfo{person}{Ben-Tzion Karsh}.} \bibinfo{year}{2010}\natexlab{}.
\newblock \showarticletitle{The technology acceptance model: its past and its future in health care}.
\newblock \bibinfo{journal}{\emph{Journal of biomedical informatics}} \bibinfo{volume}{43}, \bibinfo{number}{1} (\bibinfo{year}{2010}), \bibinfo{pages}{159--172}.
\newblock


\bibitem[\protect\citeauthoryear{Hou, Ericson, and Wang}{Hou et~al\mbox{.}}{2022}]%
        {hou2022using}
\bibfield{author}{\bibinfo{person}{Xinying Hou}, \bibinfo{person}{Barbara~Jane Ericson}, {and} \bibinfo{person}{Xu Wang}.} \bibinfo{year}{2022}\natexlab{}.
\newblock \showarticletitle{Using adaptive parsons problems to scaffold write-code problems}. In \bibinfo{booktitle}{\emph{Proceedings of the 2022 ACM Conference on International Computing Education Research-Volume 1}}. \bibinfo{pages}{15--26}.
\newblock


\bibitem[\protect\citeauthoryear{Hou, Zhao, Liu, Yang, Wang, Li, Luo, Lo, Grundy, and Wang}{Hou et~al\mbox{.}}{2024}]%
        {hou2024large}
\bibfield{author}{\bibinfo{person}{Xinyi Hou}, \bibinfo{person}{Yanjie Zhao}, \bibinfo{person}{Yue Liu}, \bibinfo{person}{Zhou Yang}, \bibinfo{person}{Kailong Wang}, \bibinfo{person}{Li Li}, \bibinfo{person}{Xiapu Luo}, \bibinfo{person}{David Lo}, \bibinfo{person}{John Grundy}, {and} \bibinfo{person}{Haoyu Wang}.} \bibinfo{year}{2024}\natexlab{}.
\newblock \showarticletitle{Large language models for software engineering: A systematic literature review}.
\newblock \bibinfo{journal}{\emph{ACM Transactions on Software Engineering and Methodology}} \bibinfo{volume}{33}, \bibinfo{number}{8} (\bibinfo{year}{2024}), \bibinfo{pages}{1--79}.
\newblock


\bibitem[\protect\citeauthoryear{Hsieh and Shannon}{Hsieh and Shannon}{2005}]%
        {hsieh2005three}
\bibfield{author}{\bibinfo{person}{Hsiu-Fang Hsieh} {and} \bibinfo{person}{Sarah~E Shannon}.} \bibinfo{year}{2005}\natexlab{}.
\newblock \showarticletitle{Three approaches to qualitative content analysis}.
\newblock \bibinfo{journal}{\emph{Qualitative health research}} \bibinfo{volume}{15}, \bibinfo{number}{9} (\bibinfo{year}{2005}), \bibinfo{pages}{1277--1288}.
\newblock


\bibitem[\protect\citeauthoryear{Jin, Lee, Shin, and Kim}{Jin et~al\mbox{.}}{2024}]%
        {jin2024teach}
\bibfield{author}{\bibinfo{person}{Hyoungwook Jin}, \bibinfo{person}{Seonghee Lee}, \bibinfo{person}{Hyungyu Shin}, {and} \bibinfo{person}{Juho Kim}.} \bibinfo{year}{2024}\natexlab{}.
\newblock \showarticletitle{Teach AI How to Code: Using Large Language Models as Teachable Agents for Programming Education}. In \bibinfo{booktitle}{\emph{Proceedings of the CHI Conference on Human Factors in Computing Systems}}. \bibinfo{pages}{1--28}.
\newblock


\bibitem[\protect\citeauthoryear{Kamin, O'Sullivan, Younger, and Deterding}{Kamin et~al\mbox{.}}{2001}]%
        {kamin2001measuring}
\bibfield{author}{\bibinfo{person}{Carol~S Kamin}, \bibinfo{person}{Patricia~S O'Sullivan}, \bibinfo{person}{Monica Younger}, {and} \bibinfo{person}{Robin Deterding}.} \bibinfo{year}{2001}\natexlab{}.
\newblock \showarticletitle{Measuring critical thinking in problem-based learning discourse}.
\newblock \bibinfo{journal}{\emph{Teaching and learning in medicine}} \bibinfo{volume}{13}, \bibinfo{number}{1} (\bibinfo{year}{2001}), \bibinfo{pages}{27--35}.
\newblock


\bibitem[\protect\citeauthoryear{Kasneci, Se{\ss}ler, K{\"u}chemann, Bannert, Dementieva, Fischer, Gasser, Groh, G{\"u}nnemann, H{\"u}llermeier, et~al\mbox{.}}{Kasneci et~al\mbox{.}}{2023}]%
        {kasneci2023chatgpt}
\bibfield{author}{\bibinfo{person}{Enkelejda Kasneci}, \bibinfo{person}{Kathrin Se{\ss}ler}, \bibinfo{person}{Stefan K{\"u}chemann}, \bibinfo{person}{Maria Bannert}, \bibinfo{person}{Daryna Dementieva}, \bibinfo{person}{Frank Fischer}, \bibinfo{person}{Urs Gasser}, \bibinfo{person}{Georg Groh}, \bibinfo{person}{Stephan G{\"u}nnemann}, \bibinfo{person}{Eyke H{\"u}llermeier}, {et~al\mbox{.}}} \bibinfo{year}{2023}\natexlab{}.
\newblock \showarticletitle{ChatGPT for good? On opportunities and challenges of large language models for education}.
\newblock \bibinfo{journal}{\emph{Learning and individual differences}}  \bibinfo{volume}{103} (\bibinfo{year}{2023}), \bibinfo{pages}{102274}.
\newblock


\bibitem[\protect\citeauthoryear{Kazemitabaar, Chow, Ma, Ericson, Weintrop, and Grossman}{Kazemitabaar et~al\mbox{.}}{2023a}]%
        {kazemitabaar2023studying}
\bibfield{author}{\bibinfo{person}{Majeed Kazemitabaar}, \bibinfo{person}{Justin Chow}, \bibinfo{person}{Carl Ka~To Ma}, \bibinfo{person}{Barbara~J Ericson}, \bibinfo{person}{David Weintrop}, {and} \bibinfo{person}{Tovi Grossman}.} \bibinfo{year}{2023}\natexlab{a}.
\newblock \showarticletitle{Studying the effect of AI code generators on supporting novice learners in introductory programming}. In \bibinfo{booktitle}{\emph{Proceedings of the 2023 CHI Conference on Human Factors in Computing Systems}}. \bibinfo{pages}{1--23}.
\newblock


\bibitem[\protect\citeauthoryear{Kazemitabaar, Hou, Henley, Ericson, Weintrop, and Grossman}{Kazemitabaar et~al\mbox{.}}{2023b}]%
        {kazemitabaar2023novices}
\bibfield{author}{\bibinfo{person}{Majeed Kazemitabaar}, \bibinfo{person}{Xinying Hou}, \bibinfo{person}{Austin Henley}, \bibinfo{person}{Barbara~Jane Ericson}, \bibinfo{person}{David Weintrop}, {and} \bibinfo{person}{Tovi Grossman}.} \bibinfo{year}{2023}\natexlab{b}.
\newblock \showarticletitle{How novices use LLM-based code generators to solve CS1 coding tasks in a self-paced learning environment}. In \bibinfo{booktitle}{\emph{Proceedings of the 23rd Koli Calling International Conference on Computing Education Research}}. \bibinfo{pages}{1--12}.
\newblock


\bibitem[\protect\citeauthoryear{Kazemitabaar, Williams, Drosos, Grossman, Henley, Negreanu, and Sarkar}{Kazemitabaar et~al\mbox{.}}{2024}]%
        {kazemitabaar2024improving}
\bibfield{author}{\bibinfo{person}{Majeed Kazemitabaar}, \bibinfo{person}{Jack Williams}, \bibinfo{person}{Ian Drosos}, \bibinfo{person}{Tovi Grossman}, \bibinfo{person}{Austin~Zachary Henley}, \bibinfo{person}{Carina Negreanu}, {and} \bibinfo{person}{Advait Sarkar}.} \bibinfo{year}{2024}\natexlab{}.
\newblock \showarticletitle{Improving steering and verification in AI-assisted data analysis with interactive task decomposition}. In \bibinfo{booktitle}{\emph{Proceedings of the 37th Annual ACM Symposium on User Interface Software and Technology}}. \bibinfo{pages}{1--19}.
\newblock


\bibitem[\protect\citeauthoryear{Keen and Mammen}{Keen and Mammen}{2015}]%
        {keen2015program}
\bibfield{author}{\bibinfo{person}{Aaron Keen} {and} \bibinfo{person}{Kurt Mammen}.} \bibinfo{year}{2015}\natexlab{}.
\newblock \showarticletitle{Program decomposition and complexity in CS1}. In \bibinfo{booktitle}{\emph{Proceedings of the 46th ACM technical symposium on computer science education}}. \bibinfo{pages}{48--53}.
\newblock


\bibitem[\protect\citeauthoryear{Kim and Hannafin}{Kim and Hannafin}{2011}]%
        {kim2011scaffolding}
\bibfield{author}{\bibinfo{person}{Minchi~C Kim} {and} \bibinfo{person}{Michael~J Hannafin}.} \bibinfo{year}{2011}\natexlab{}.
\newblock \showarticletitle{Scaffolding problem solving in technology-enhanced learning environments (TELEs): Bridging research and theory with practice}.
\newblock \bibinfo{journal}{\emph{Computers \& Education}} \bibinfo{volume}{56}, \bibinfo{number}{2} (\bibinfo{year}{2011}), \bibinfo{pages}{403--417}.
\newblock


\bibitem[\protect\citeauthoryear{Kinnunen and Malmi}{Kinnunen and Malmi}{2006}]%
        {kinnunen2006students}
\bibfield{author}{\bibinfo{person}{P{\"a}ivi Kinnunen} {and} \bibinfo{person}{Lauri Malmi}.} \bibinfo{year}{2006}\natexlab{}.
\newblock \showarticletitle{Why students drop out CS1 course?}. In \bibinfo{booktitle}{\emph{Proceedings of the second international workshop on Computing education research}}. \bibinfo{pages}{97--108}.
\newblock


\bibitem[\protect\citeauthoryear{Leinonen, Hellas, Sarsa, Reeves, Denny, Prather, and Becker}{Leinonen et~al\mbox{.}}{2023}]%
        {leinonen2023using}
\bibfield{author}{\bibinfo{person}{Juho Leinonen}, \bibinfo{person}{Arto Hellas}, \bibinfo{person}{Sami Sarsa}, \bibinfo{person}{Brent Reeves}, \bibinfo{person}{Paul Denny}, \bibinfo{person}{James Prather}, {and} \bibinfo{person}{Brett~A Becker}.} \bibinfo{year}{2023}\natexlab{}.
\newblock \showarticletitle{Using large language models to enhance programming error messages}. In \bibinfo{booktitle}{\emph{Proceedings of the 54th ACM Technical Symposium on Computer Science Education V. 1}}. \bibinfo{pages}{563--569}.
\newblock


\bibitem[\protect\citeauthoryear{Linn and Dalbey}{Linn and Dalbey}{1985}]%
        {linn1985cognitive}
\bibfield{author}{\bibinfo{person}{Marcia~C Linn} {and} \bibinfo{person}{John Dalbey}.} \bibinfo{year}{1985}\natexlab{}.
\newblock \showarticletitle{Cognitive consequences of programming instruction: Instruction, access, and ability}.
\newblock \bibinfo{journal}{\emph{Educational Psychologist}} \bibinfo{volume}{20}, \bibinfo{number}{4} (\bibinfo{year}{1985}), \bibinfo{pages}{191--206}.
\newblock


\bibitem[\protect\citeauthoryear{Liu, Xia, Wang, and Zhang}{Liu et~al\mbox{.}}{2024}]%
        {liu2024your}
\bibfield{author}{\bibinfo{person}{Jiawei Liu}, \bibinfo{person}{Chunqiu~Steven Xia}, \bibinfo{person}{Yuyao Wang}, {and} \bibinfo{person}{Lingming Zhang}.} \bibinfo{year}{2024}\natexlab{}.
\newblock \showarticletitle{Is your code generated by chatgpt really correct? rigorous evaluation of large language models for code generation}.
\newblock \bibinfo{journal}{\emph{Advances in Neural Information Processing Systems}}  \bibinfo{volume}{36} (\bibinfo{year}{2024}).
\newblock


\bibitem[\protect\citeauthoryear{Liu, Sarkar, Negreanu, Zorn, Williams, Toronto, and Gordon}{Liu et~al\mbox{.}}{2023}]%
        {liu2023wants}
\bibfield{author}{\bibinfo{person}{Michael~Xieyang Liu}, \bibinfo{person}{Advait Sarkar}, \bibinfo{person}{Carina Negreanu}, \bibinfo{person}{Benjamin Zorn}, \bibinfo{person}{Jack Williams}, \bibinfo{person}{Neil Toronto}, {and} \bibinfo{person}{Andrew~D Gordon}.} \bibinfo{year}{2023}\natexlab{}.
\newblock \showarticletitle{“What it wants me to say”: Bridging the abstraction gap between end-user programmers and code-generating large language models}. In \bibinfo{booktitle}{\emph{Proceedings of the 2023 CHI Conference on Human Factors in Computing Systems}}. \bibinfo{pages}{1--31}.
\newblock


\bibitem[\protect\citeauthoryear{Liu and Stasko}{Liu and Stasko}{2010}]%
        {liu2010mental}
\bibfield{author}{\bibinfo{person}{Zhicheng Liu} {and} \bibinfo{person}{John Stasko}.} \bibinfo{year}{2010}\natexlab{}.
\newblock \showarticletitle{Mental models, visual reasoning and interaction in information visualization: A top-down perspective}.
\newblock \bibinfo{journal}{\emph{IEEE transactions on visualization and computer graphics}} \bibinfo{volume}{16}, \bibinfo{number}{6} (\bibinfo{year}{2010}), \bibinfo{pages}{999--1008}.
\newblock


\bibitem[\protect\citeauthoryear{Ma, Chen, Wang, Zheng, Peng, Yin, and Ma}{Ma et~al\mbox{.}}{2024a}]%
        {ma2024towards}
\bibfield{author}{\bibinfo{person}{Shuai Ma}, \bibinfo{person}{Qiaoyi Chen}, \bibinfo{person}{Xinru Wang}, \bibinfo{person}{Chengbo Zheng}, \bibinfo{person}{Zhenhui Peng}, \bibinfo{person}{Ming Yin}, {and} \bibinfo{person}{Xiaojuan Ma}.} \bibinfo{year}{2024}\natexlab{a}.
\newblock \showarticletitle{Towards human-ai deliberation: Design and evaluation of llm-empowered deliberative ai for ai-assisted decision-making}.
\newblock \bibinfo{journal}{\emph{arXiv preprint arXiv:2403.16812}} (\bibinfo{year}{2024}).
\newblock


\bibitem[\protect\citeauthoryear{Ma, Lei, Wang, Zheng, Shi, Yin, and Ma}{Ma et~al\mbox{.}}{2023}]%
        {ma2023should}
\bibfield{author}{\bibinfo{person}{Shuai Ma}, \bibinfo{person}{Ying Lei}, \bibinfo{person}{Xinru Wang}, \bibinfo{person}{Chengbo Zheng}, \bibinfo{person}{Chuhan Shi}, \bibinfo{person}{Ming Yin}, {and} \bibinfo{person}{Xiaojuan Ma}.} \bibinfo{year}{2023}\natexlab{}.
\newblock \showarticletitle{Who Should I Trust: AI or Myself? Leveraging Human and AI Correctness Likelihood to Promote Appropriate Trust in AI-Assisted Decision-Making}. In \bibinfo{booktitle}{\emph{Proceedings of the 2023 CHI Conference on Human Factors in Computing Systems}}. \bibinfo{pages}{1--19}.
\newblock


\bibitem[\protect\citeauthoryear{Ma, Wang, Lei, Shi, Yin, and Ma}{Ma et~al\mbox{.}}{2024b}]%
        {ma2024you}
\bibfield{author}{\bibinfo{person}{Shuai Ma}, \bibinfo{person}{Xinru Wang}, \bibinfo{person}{Ying Lei}, \bibinfo{person}{Chuhan Shi}, \bibinfo{person}{Ming Yin}, {and} \bibinfo{person}{Xiaojuan Ma}.} \bibinfo{year}{2024}\natexlab{b}.
\newblock \showarticletitle{" Are You Really Sure?" Understanding the Effects of Human Self-Confidence Calibration in AI-Assisted Decision Making}.
\newblock \bibinfo{journal}{\emph{arXiv preprint arXiv:2403.09552}} (\bibinfo{year}{2024}).
\newblock


\bibitem[\protect\citeauthoryear{Ma, Wei, Tian, Fan, Zhang, Shen, Lin, Huang, M{\v{e}}ch, Samaras, et~al\mbox{.}}{Ma et~al\mbox{.}}{2019}]%
        {ma2019smarteye}
\bibfield{author}{\bibinfo{person}{Shuai Ma}, \bibinfo{person}{Zijun Wei}, \bibinfo{person}{Feng Tian}, \bibinfo{person}{Xiangmin Fan}, \bibinfo{person}{Jianming Zhang}, \bibinfo{person}{Xiaohui Shen}, \bibinfo{person}{Zhe Lin}, \bibinfo{person}{Jin Huang}, \bibinfo{person}{Radom{\'\i}r M{\v{e}}ch}, \bibinfo{person}{Dimitris Samaras}, {et~al\mbox{.}}} \bibinfo{year}{2019}\natexlab{}.
\newblock \showarticletitle{SmartEye: assisting instant photo taking via integrating user preference with deep view proposal network}. In \bibinfo{booktitle}{\emph{Proceedings of the 2019 CHI conference on human factors in computing systems}}. \bibinfo{pages}{1--12}.
\newblock


\bibitem[\protect\citeauthoryear{Ma, Zhou, Nie, and Ma}{Ma et~al\mbox{.}}{2022}]%
        {ma2022glancee}
\bibfield{author}{\bibinfo{person}{Shuai Ma}, \bibinfo{person}{Taichang Zhou}, \bibinfo{person}{Fei Nie}, {and} \bibinfo{person}{Xiaojuan Ma}.} \bibinfo{year}{2022}\natexlab{}.
\newblock \showarticletitle{Glancee: An Adaptable System for Instructors to Grasp Student Learning Status in Synchronous Online Classes}. In \bibinfo{booktitle}{\emph{CHI Conference on Human Factors in Computing Systems}}. \bibinfo{pages}{1--25}.
\newblock


\bibitem[\protect\citeauthoryear{MacNeil, Tran, Hellas, Kim, Sarsa, Denny, Bernstein, and Leinonen}{MacNeil et~al\mbox{.}}{2023}]%
        {macneil2023experiences}
\bibfield{author}{\bibinfo{person}{Stephen MacNeil}, \bibinfo{person}{Andrew Tran}, \bibinfo{person}{Arto Hellas}, \bibinfo{person}{Joanne Kim}, \bibinfo{person}{Sami Sarsa}, \bibinfo{person}{Paul Denny}, \bibinfo{person}{Seth Bernstein}, {and} \bibinfo{person}{Juho Leinonen}.} \bibinfo{year}{2023}\natexlab{}.
\newblock \showarticletitle{Experiences from using code explanations generated by large language models in a web software development e-book}. In \bibinfo{booktitle}{\emph{Proceedings of the 54th ACM Technical Symposium on Computer Science Education V. 1}}. \bibinfo{pages}{931--937}.
\newblock


\bibitem[\protect\citeauthoryear{McCracken, Almstrum, Diaz, Guzdial, Hagan, Kolikant, Laxer, Thomas, Utting, and Wilusz}{McCracken et~al\mbox{.}}{2001}]%
        {mccracken2001multi}
\bibfield{author}{\bibinfo{person}{Michael McCracken}, \bibinfo{person}{Vicki Almstrum}, \bibinfo{person}{Danny Diaz}, \bibinfo{person}{Mark Guzdial}, \bibinfo{person}{Dianne Hagan}, \bibinfo{person}{Yifat Ben-David Kolikant}, \bibinfo{person}{Cary Laxer}, \bibinfo{person}{Lynda Thomas}, \bibinfo{person}{Ian Utting}, {and} \bibinfo{person}{Tadeusz Wilusz}.} \bibinfo{year}{2001}\natexlab{}.
\newblock \showarticletitle{A multi-national, multi-institutional study of assessment of programming skills of first-year CS students}.
\newblock In \bibinfo{booktitle}{\emph{Working group reports from ITiCSE on Innovation and technology in computer science education}}. \bibinfo{pages}{125--180}.
\newblock


\bibitem[\protect\citeauthoryear{McDougall, Curry, and De~Bruijn}{McDougall et~al\mbox{.}}{2001}]%
        {mcdougall2001effects}
\bibfield{author}{\bibinfo{person}{Sin{\'e}~JP McDougall}, \bibinfo{person}{Martin~B Curry}, {and} \bibinfo{person}{Oscar De~Bruijn}.} \bibinfo{year}{2001}\natexlab{}.
\newblock \showarticletitle{The effects of visual information on users' mental models: An evaluation of Pathfinder analysis as a measure of icon usability}.
\newblock \bibinfo{journal}{\emph{International journal of cognitive ergonomics}} \bibinfo{volume}{5}, \bibinfo{number}{1} (\bibinfo{year}{2001}), \bibinfo{pages}{59--84}.
\newblock


\bibitem[\protect\citeauthoryear{Pea}{Pea}{1987}]%
        {pea1987logo}
\bibfield{author}{\bibinfo{person}{Roy~D Pea}.} \bibinfo{year}{1987}\natexlab{}.
\newblock \showarticletitle{Logo programming and problem solving}.
\newblock  (\bibinfo{year}{1987}).
\newblock


\bibitem[\protect\citeauthoryear{Pea}{Pea}{2018}]%
        {pea2018social}
\bibfield{author}{\bibinfo{person}{Roy~D Pea}.} \bibinfo{year}{2018}\natexlab{}.
\newblock \showarticletitle{The social and technological dimensions of scaffolding and related theoretical concepts for learning, education, and human activity}.
\newblock In \bibinfo{booktitle}{\emph{Scaffolding}}. \bibinfo{publisher}{Psychology Press}, \bibinfo{pages}{423--451}.
\newblock


\bibitem[\protect\citeauthoryear{Pea and Kurland}{Pea and Kurland}{1984}]%
        {pea1984cognitive}
\bibfield{author}{\bibinfo{person}{Roy~D Pea} {and} \bibinfo{person}{D~Midian Kurland}.} \bibinfo{year}{1984}\natexlab{}.
\newblock \showarticletitle{On the cognitive effects of learning computer programming}.
\newblock \bibinfo{journal}{\emph{New ideas in psychology}} \bibinfo{volume}{2}, \bibinfo{number}{2} (\bibinfo{year}{1984}), \bibinfo{pages}{137--168}.
\newblock


\bibitem[\protect\citeauthoryear{Pearce, Nakazawa, and Heggen}{Pearce et~al\mbox{.}}{2015}]%
        {pearce2015improving}
\bibfield{author}{\bibinfo{person}{Janice~L Pearce}, \bibinfo{person}{Mario Nakazawa}, {and} \bibinfo{person}{Scott Heggen}.} \bibinfo{year}{2015}\natexlab{}.
\newblock \showarticletitle{Improving problem decomposition ability in CS1 through explicit guided inquiry-based instruction}.
\newblock \bibinfo{journal}{\emph{J. Comput. Sci. Coll}} \bibinfo{volume}{31}, \bibinfo{number}{2} (\bibinfo{year}{2015}), \bibinfo{pages}{135--144}.
\newblock


\bibitem[\protect\citeauthoryear{Phung, Cambronero, Gulwani, Kohn, Majumdar, Singla, and Soares}{Phung et~al\mbox{.}}{2023}]%
        {phung2023generating}
\bibfield{author}{\bibinfo{person}{Tung Phung}, \bibinfo{person}{Jos{\'e} Cambronero}, \bibinfo{person}{Sumit Gulwani}, \bibinfo{person}{Tobias Kohn}, \bibinfo{person}{Rupak Majumdar}, \bibinfo{person}{Adish Singla}, {and} \bibinfo{person}{Gustavo Soares}.} \bibinfo{year}{2023}\natexlab{}.
\newblock \showarticletitle{Generating high-precision feedback for programming syntax errors using large language models}.
\newblock \bibinfo{journal}{\emph{arXiv preprint arXiv:2302.04662}} (\bibinfo{year}{2023}).
\newblock


\bibitem[\protect\citeauthoryear{Piech, Sahami, Huang, and Guibas}{Piech et~al\mbox{.}}{2015}]%
        {piech2015autonomously}
\bibfield{author}{\bibinfo{person}{Chris Piech}, \bibinfo{person}{Mehran Sahami}, \bibinfo{person}{Jonathan Huang}, {and} \bibinfo{person}{Leonidas Guibas}.} \bibinfo{year}{2015}\natexlab{}.
\newblock \showarticletitle{Autonomously generating hints by inferring problem solving policies}. In \bibinfo{booktitle}{\emph{Proceedings of the second (2015) acm conference on learning@ scale}}. \bibinfo{pages}{195--204}.
\newblock


\bibitem[\protect\citeauthoryear{Pitterson, Brown, Pascoe, and Fisher}{Pitterson et~al\mbox{.}}{2016}]%
        {pitterson2016measuring}
\bibfield{author}{\bibinfo{person}{Nicole~P Pitterson}, \bibinfo{person}{Shane Brown}, \bibinfo{person}{Jason Pascoe}, {and} \bibinfo{person}{Kathleen~Quardokus Fisher}.} \bibinfo{year}{2016}\natexlab{}.
\newblock \showarticletitle{Measuring cognitive engagement through interactive, constructive, active and passive learning activities}. In \bibinfo{booktitle}{\emph{2016 IEEE Frontiers in Education Conference (FIE)}}. IEEE, \bibinfo{pages}{1--6}.
\newblock


\bibitem[\protect\citeauthoryear{Prather, Reeves, Denny, Becker, Leinonen, Luxton-Reilly, Powell, Finnie-Ansley, and Santos}{Prather et~al\mbox{.}}{2023}]%
        {prather2023s}
\bibfield{author}{\bibinfo{person}{James Prather}, \bibinfo{person}{Brent~N Reeves}, \bibinfo{person}{Paul Denny}, \bibinfo{person}{Brett~A Becker}, \bibinfo{person}{Juho Leinonen}, \bibinfo{person}{Andrew Luxton-Reilly}, \bibinfo{person}{Garrett Powell}, \bibinfo{person}{James Finnie-Ansley}, {and} \bibinfo{person}{Eddie~Antonio Santos}.} \bibinfo{year}{2023}\natexlab{}.
\newblock \showarticletitle{“It’s Weird That it Knows What I Want”: Usability and Interactions with Copilot for Novice Programmers}.
\newblock \bibinfo{journal}{\emph{ACM Transactions on Computer-Human Interaction}} \bibinfo{volume}{31}, \bibinfo{number}{1} (\bibinfo{year}{2023}), \bibinfo{pages}{1--31}.
\newblock


\bibitem[\protect\citeauthoryear{Qian and Lehman}{Qian and Lehman}{2017}]%
        {qian2017students}
\bibfield{author}{\bibinfo{person}{Yizhou Qian} {and} \bibinfo{person}{James Lehman}.} \bibinfo{year}{2017}\natexlab{}.
\newblock \showarticletitle{Students’ misconceptions and other difficulties in introductory programming: A literature review}.
\newblock \bibinfo{journal}{\emph{ACM Transactions on Computing Education (TOCE)}} \bibinfo{volume}{18}, \bibinfo{number}{1} (\bibinfo{year}{2017}), \bibinfo{pages}{1--24}.
\newblock


\bibitem[\protect\citeauthoryear{Rivers and Koedinger}{Rivers and Koedinger}{2017}]%
        {rivers2017data}
\bibfield{author}{\bibinfo{person}{Kelly Rivers} {and} \bibinfo{person}{Kenneth~R Koedinger}.} \bibinfo{year}{2017}\natexlab{}.
\newblock \showarticletitle{Data-driven hint generation in vast solution spaces: a self-improving python programming tutor}.
\newblock \bibinfo{journal}{\emph{International Journal of Artificial Intelligence in Education}}  \bibinfo{volume}{27} (\bibinfo{year}{2017}), \bibinfo{pages}{37--64}.
\newblock


\bibitem[\protect\citeauthoryear{Roest, Keuning, and Jeuring}{Roest et~al\mbox{.}}{2024}]%
        {roest2024next}
\bibfield{author}{\bibinfo{person}{Lianne Roest}, \bibinfo{person}{Hieke Keuning}, {and} \bibinfo{person}{Johan Jeuring}.} \bibinfo{year}{2024}\natexlab{}.
\newblock \showarticletitle{Next-Step Hint Generation for Introductory Programming Using Large Language Models}. In \bibinfo{booktitle}{\emph{Proceedings of the 26th Australasian Computing Education Conference}}. \bibinfo{pages}{144--153}.
\newblock


\bibitem[\protect\citeauthoryear{Sarsa, Denny, Hellas, and Leinonen}{Sarsa et~al\mbox{.}}{2022}]%
        {sarsa2022automatic}
\bibfield{author}{\bibinfo{person}{Sami Sarsa}, \bibinfo{person}{Paul Denny}, \bibinfo{person}{Arto Hellas}, {and} \bibinfo{person}{Juho Leinonen}.} \bibinfo{year}{2022}\natexlab{}.
\newblock \showarticletitle{Automatic generation of programming exercises and code explanations using large language models}. In \bibinfo{booktitle}{\emph{Proceedings of the 2022 ACM Conference on International Computing Education Research-Volume 1}}. \bibinfo{pages}{27--43}.
\newblock


\bibitem[\protect\citeauthoryear{Savelka, Agarwal, An, Bogart, and Sakr}{Savelka et~al\mbox{.}}{2023}]%
        {savelka2023thrilled}
\bibfield{author}{\bibinfo{person}{Jaromir Savelka}, \bibinfo{person}{Arav Agarwal}, \bibinfo{person}{Marshall An}, \bibinfo{person}{Chris Bogart}, {and} \bibinfo{person}{Majd Sakr}.} \bibinfo{year}{2023}\natexlab{}.
\newblock \showarticletitle{Thrilled by your progress! large language models (gpt-4) no longer struggle to pass assessments in higher education programming courses}. In \bibinfo{booktitle}{\emph{Proceedings of the 2023 ACM Conference on International Computing Education Research-Volume 1}}. \bibinfo{pages}{78--92}.
\newblock


\bibitem[\protect\citeauthoryear{Schmidt, Loyens, Van~Gog, and Paas}{Schmidt et~al\mbox{.}}{2007}]%
        {schmidt2007problem}
\bibfield{author}{\bibinfo{person}{Henk~G Schmidt}, \bibinfo{person}{Sofie~MM Loyens}, \bibinfo{person}{Tamara Van~Gog}, {and} \bibinfo{person}{Fred Paas}.} \bibinfo{year}{2007}\natexlab{}.
\newblock \showarticletitle{Problem-based learning is compatible with human cognitive architecture: Commentary on Kirschner, Sweller, and}.
\newblock \bibinfo{journal}{\emph{Educational psychologist}} \bibinfo{volume}{42}, \bibinfo{number}{2} (\bibinfo{year}{2007}), \bibinfo{pages}{91--97}.
\newblock


\bibitem[\protect\citeauthoryear{Sheese, Liffiton, Savelka, and Denny}{Sheese et~al\mbox{.}}{2024}]%
        {sheese2024patterns}
\bibfield{author}{\bibinfo{person}{Brad Sheese}, \bibinfo{person}{Mark Liffiton}, \bibinfo{person}{Jaromir Savelka}, {and} \bibinfo{person}{Paul Denny}.} \bibinfo{year}{2024}\natexlab{}.
\newblock \showarticletitle{Patterns of student help-seeking when using a large language model-powered programming assistant}. In \bibinfo{booktitle}{\emph{Proceedings of the 26th Australasian Computing Education Conference}}. \bibinfo{pages}{49--57}.
\newblock


\bibitem[\protect\citeauthoryear{Singh, Gulwani, and Solar-Lezama}{Singh et~al\mbox{.}}{2013}]%
        {singh2013automated}
\bibfield{author}{\bibinfo{person}{Rishabh Singh}, \bibinfo{person}{Sumit Gulwani}, {and} \bibinfo{person}{Armando Solar-Lezama}.} \bibinfo{year}{2013}\natexlab{}.
\newblock \showarticletitle{Automated feedback generation for introductory programming assignments}. In \bibinfo{booktitle}{\emph{Proceedings of the 34th ACM SIGPLAN conference on Programming language design and implementation}}. \bibinfo{pages}{15--26}.
\newblock


\bibitem[\protect\citeauthoryear{Skiena}{Skiena}{1998}]%
        {skiena1998algorithm}
\bibfield{author}{\bibinfo{person}{Steven~S Skiena}.} \bibinfo{year}{1998}\natexlab{}.
\newblock \bibinfo{booktitle}{\emph{The algorithm design manual}}. Vol.~\bibinfo{volume}{2}.
\newblock \bibinfo{publisher}{Springer}.
\newblock


\bibitem[\protect\citeauthoryear{Smetsers-Weeda and Smetsers}{Smetsers-Weeda and Smetsers}{2017}]%
        {smetsers2017problem}
\bibfield{author}{\bibinfo{person}{Renske Smetsers-Weeda} {and} \bibinfo{person}{Sjaak Smetsers}.} \bibinfo{year}{2017}\natexlab{}.
\newblock \showarticletitle{Problem solving and algorithmic development with flowcharts}. In \bibinfo{booktitle}{\emph{Proceedings of the 12th Workshop on Primary and Secondary Computing Education}}. \bibinfo{pages}{25--34}.
\newblock


\bibitem[\protect\citeauthoryear{Sooriamurthi}{Sooriamurthi}{2009}]%
        {sooriamurthi2009introducing}
\bibfield{author}{\bibinfo{person}{Raja Sooriamurthi}.} \bibinfo{year}{2009}\natexlab{}.
\newblock \showarticletitle{Introducing abstraction and decomposition to novice programmers}.
\newblock \bibinfo{journal}{\emph{ACM SIGCSE Bulletin}} \bibinfo{volume}{41}, \bibinfo{number}{3} (\bibinfo{year}{2009}), \bibinfo{pages}{196--200}.
\newblock


\bibitem[\protect\citeauthoryear{Sykes}{Sykes}{2010}]%
        {sykes2010design}
\bibfield{author}{\bibinfo{person}{Edward~R Sykes}.} \bibinfo{year}{2010}\natexlab{}.
\newblock \showarticletitle{Design, Development and Evaluation of the Java Intelligent Tutoring System.}
\newblock \bibinfo{journal}{\emph{Technology, Instruction, Cognition \& Learning}} \bibinfo{volume}{8}, \bibinfo{number}{1} (\bibinfo{year}{2010}).
\newblock


\bibitem[\protect\citeauthoryear{Tobias and Duffy}{Tobias and Duffy}{2009}]%
        {tobias2009constructivist}
\bibfield{author}{\bibinfo{person}{Sigmund Tobias} {and} \bibinfo{person}{Thomas~M Duffy}.} \bibinfo{year}{2009}\natexlab{}.
\newblock \showarticletitle{Constructivist instruction}.
\newblock \bibinfo{journal}{\emph{Success or failure}} (\bibinfo{year}{2009}).
\newblock


\bibitem[\protect\citeauthoryear{Tsai}{Tsai}{2019}]%
        {tsai2019improving}
\bibfield{author}{\bibinfo{person}{Chun-Yen Tsai}.} \bibinfo{year}{2019}\natexlab{}.
\newblock \showarticletitle{Improving students' understanding of basic programming concepts through visual programming language: The role of self-efficacy}.
\newblock \bibinfo{journal}{\emph{Computers in Human Behavior}}  \bibinfo{volume}{95} (\bibinfo{year}{2019}), \bibinfo{pages}{224--232}.
\newblock


\bibitem[\protect\citeauthoryear{Van~de Pol, Volman, and Beishuizen}{Van~de Pol et~al\mbox{.}}{2010}]%
        {van2010scaffolding}
\bibfield{author}{\bibinfo{person}{Janneke Van~de Pol}, \bibinfo{person}{Monique Volman}, {and} \bibinfo{person}{Jos Beishuizen}.} \bibinfo{year}{2010}\natexlab{}.
\newblock \showarticletitle{Scaffolding in teacher--student interaction: A decade of research}.
\newblock \bibinfo{journal}{\emph{Educational psychology review}}  \bibinfo{volume}{22} (\bibinfo{year}{2010}), \bibinfo{pages}{271--296}.
\newblock


\bibitem[\protect\citeauthoryear{VanLehn}{VanLehn}{2011}]%
        {vanlehn2011relative}
\bibfield{author}{\bibinfo{person}{Kurt VanLehn}.} \bibinfo{year}{2011}\natexlab{}.
\newblock \showarticletitle{The relative effectiveness of human tutoring, intelligent tutoring systems, and other tutoring systems}.
\newblock \bibinfo{journal}{\emph{Educational psychologist}} \bibinfo{volume}{46}, \bibinfo{number}{4} (\bibinfo{year}{2011}), \bibinfo{pages}{197--221}.
\newblock


\bibitem[\protect\citeauthoryear{Wang, Le~Meur, Bobbadi, Akram, Barnes, Martens, and Price}{Wang et~al\mbox{.}}{2022}]%
        {wang2022exploring}
\bibfield{author}{\bibinfo{person}{Wengran Wang}, \bibinfo{person}{Audrey Le~Meur}, \bibinfo{person}{Mahesh Bobbadi}, \bibinfo{person}{Bita Akram}, \bibinfo{person}{Tiffany Barnes}, \bibinfo{person}{Chris Martens}, {and} \bibinfo{person}{Thomas Price}.} \bibinfo{year}{2022}\natexlab{}.
\newblock \showarticletitle{Exploring design choices to support novices' example use during creative open-ended programming}. In \bibinfo{booktitle}{\emph{Proceedings of the 53rd ACM Technical Symposium on Computer Science Education-Volume 1}}. \bibinfo{pages}{619--625}.
\newblock


\bibitem[\protect\citeauthoryear{Wei, Wang, Schuurmans, Bosma, Xia, Chi, Le, Zhou, et~al\mbox{.}}{Wei et~al\mbox{.}}{2022}]%
        {wei2022chain}
\bibfield{author}{\bibinfo{person}{Jason Wei}, \bibinfo{person}{Xuezhi Wang}, \bibinfo{person}{Dale Schuurmans}, \bibinfo{person}{Maarten Bosma}, \bibinfo{person}{Fei Xia}, \bibinfo{person}{Ed Chi}, \bibinfo{person}{Quoc~V Le}, \bibinfo{person}{Denny Zhou}, {et~al\mbox{.}}} \bibinfo{year}{2022}\natexlab{}.
\newblock \showarticletitle{Chain-of-thought prompting elicits reasoning in large language models}.
\newblock \bibinfo{journal}{\emph{Advances in neural information processing systems}}  \bibinfo{volume}{35} (\bibinfo{year}{2022}), \bibinfo{pages}{24824--24837}.
\newblock


\bibitem[\protect\citeauthoryear{Wiedenbeck, Labelle, and Kain}{Wiedenbeck et~al\mbox{.}}{2004}]%
        {wiedenbeck2004factors}
\bibfield{author}{\bibinfo{person}{Susan Wiedenbeck}, \bibinfo{person}{Deborah Labelle}, {and} \bibinfo{person}{Vennila~NR Kain}.} \bibinfo{year}{2004}\natexlab{}.
\newblock \showarticletitle{Factors affecting course outcomes in introductory programming.}. In \bibinfo{booktitle}{\emph{PPIG}}. \bibinfo{pages}{11}.
\newblock


\bibitem[\protect\citeauthoryear{Wing}{Wing}{2006}]%
        {wing2006computational}
\bibfield{author}{\bibinfo{person}{Jeannette~M Wing}.} \bibinfo{year}{2006}\natexlab{}.
\newblock \showarticletitle{Computational thinking}.
\newblock \bibinfo{journal}{\emph{Commun. ACM}} \bibinfo{volume}{49}, \bibinfo{number}{3} (\bibinfo{year}{2006}), \bibinfo{pages}{33--35}.
\newblock


\bibitem[\protect\citeauthoryear{Wood, Bruner, and Ross}{Wood et~al\mbox{.}}{1976}]%
        {wood1976role}
\bibfield{author}{\bibinfo{person}{David Wood}, \bibinfo{person}{Jerome~S Bruner}, {and} \bibinfo{person}{Gail Ross}.} \bibinfo{year}{1976}\natexlab{}.
\newblock \showarticletitle{The role of tutoring in problem solving}.
\newblock \bibinfo{journal}{\emph{Journal of child psychology and psychiatry}} \bibinfo{volume}{17}, \bibinfo{number}{2} (\bibinfo{year}{1976}), \bibinfo{pages}{89--100}.
\newblock


\bibitem[\protect\citeauthoryear{Ya{\u{g}}c{\i}}{Ya{\u{g}}c{\i}}{2019}]%
        {yaugci2019valid}
\bibfield{author}{\bibinfo{person}{Mustafa Ya{\u{g}}c{\i}}.} \bibinfo{year}{2019}\natexlab{}.
\newblock \showarticletitle{A valid and reliable tool for examining computational thinking skills}.
\newblock \bibinfo{journal}{\emph{Education and Information Technologies}} \bibinfo{volume}{24}, \bibinfo{number}{1} (\bibinfo{year}{2019}), \bibinfo{pages}{929--951}.
\newblock


\bibitem[\protect\citeauthoryear{Yildiz~Durak}{Yildiz~Durak}{2018}]%
        {yildiz2018digital}
\bibfield{author}{\bibinfo{person}{Hatice Yildiz~Durak}.} \bibinfo{year}{2018}\natexlab{}.
\newblock \showarticletitle{Digital story design activities used for teaching programming effect on learning of programming concepts, programming self-efficacy, and participation and analysis of student experiences}.
\newblock \bibinfo{journal}{\emph{Journal of Computer Assisted Learning}} \bibinfo{volume}{34}, \bibinfo{number}{6} (\bibinfo{year}{2018}), \bibinfo{pages}{740--752}.
\newblock


\bibitem[\protect\citeauthoryear{Zhou, Chai, Jong, and Xiong}{Zhou et~al\mbox{.}}{2021}]%
        {zhou2021does}
\bibfield{author}{\bibinfo{person}{Xiaohua Zhou}, \bibinfo{person}{Ching~Sing Chai}, \bibinfo{person}{Morris Siu-Yung Jong}, {and} \bibinfo{person}{Xi~Bei Xiong}.} \bibinfo{year}{2021}\natexlab{}.
\newblock \showarticletitle{Does relatedness matter for online self-regulated learning to promote perceived learning gains and satisfaction?}
\newblock \bibinfo{journal}{\emph{The Asia-Pacific Education Researcher}} \bibinfo{volume}{30}, \bibinfo{number}{3} (\bibinfo{year}{2021}), \bibinfo{pages}{205--215}.
\newblock


\bibitem[\protect\citeauthoryear{Zimmerman}{Zimmerman}{2013}]%
        {zimmerman2013theories}
\bibfield{author}{\bibinfo{person}{Barry~J Zimmerman}.} \bibinfo{year}{2013}\natexlab{}.
\newblock \showarticletitle{Theories of self-regulated learning and academic achievement: An overview and analysis}.
\newblock \bibinfo{journal}{\emph{Self-regulated learning and academic achievement}} (\bibinfo{year}{2013}), \bibinfo{pages}{1--36}.
\newblock


\end{thebibliography}

\end{document}